\documentclass[12pt]{iopart}

\usepackage{graphicx} 
\usepackage{dcolumn}  
\usepackage{bm}       
\usepackage{hyperref} 
\usepackage{soul,xcolor}	
\usepackage{cancel}	

\expandafter\let\csname equation*\endcsname\relax
\expandafter\let\csname endequation*\endcsname\relax
\usepackage{amssymb}
\usepackage{amsmath}

\usepackage[mathb]{mathabx}    

\usepackage{multirow}     
\usepackage{subfig}

\allowdisplaybreaks

\newcommand{\beq}{\begin{equation}}
\newcommand{\eeq}{\end{equation}}
\newcommand{\eq}{\end{equation}}

\newcommand{\ii}{{\rm i}}
\newcommand{\dd}{{\rm d}}

\newcommand{\bC}{{\mathbb{C}}}

\newcommand{\bR}{{\mathbb{R}}}

\newcommand{\secular}[1]{\left( #1 \right)_{\mathrm{sec}}}

\newcommand{\A}[1]{\secular{\dd #1/\dd t}}
\newcommand{\B}[1]{\secular{\frac{\dd #1}{\dd t}}}

\newcommand{\reff}[1]{\mbox{Figure\hspace{2pt}\ref{#1}}}

\newcommand{\refe}[1]{(\ref{#1})}

\newcommand{\refsec}[1]{Section \ref{#1}}

\newcommand{\ecc}{\ensuremath{\varepsilon}}		
\newcommand{\incl}{\ensuremath{\iota}}			

\newcommand{\pp}{\mathcal{P}}

\newcommand{\degr}{^{\circ}}

\begin{document}
\setstcolor{red}	
%
%
\title{Relativistic Positioning System in Perturbed Space-time}
\newcommand{\affLJ} {%
Faculty of Mathematics and Physics, 
University of Ljubljana, 
Jadranska ulica 19, 
1000 Ljubljana, Slovenia}
\newcommand{\affPA} {%
INAF, 
Astronomical Observatory of Padova, 
Vicolo Osservatorio 5, 
35122 Padova, Italy}
\author{
  Uro\v s Kosti\' c$^{1,2}$,
  Martin Horvat $^1$,
  Andreja Gomboc$^1$
}
\ead{uros.kostic@fmf.uni-lj.si}

\address{$^1$\affLJ}
\address{$^2$\affPA}
\date{\today}
\begin{abstract}
We present a variant of a Global Navigation Satellite System called a Relativistic Positioning System (RPS), which is based on emission coordinates. We modelled the RPS dynamics in a space-time around Earth, described by a perturbed Schwarzschild metric, where we included the perturbations due to Earth multipoles (up to the 6th), the Moon, the Sun, Venus, Jupiter, solid tide, ocean tide, and Kerr rotation effect. The exchange of signals between the satellites and a user was calculated using a ray-tracing method in the Schwarzschild space-time. We find that positioning in a perturbed space-time is feasible and is highly accurate already with standard numerical procedures: the positioning algorithms used to transform between the emission and the Schwarzschild coordinates of the user are very accurate and time efficient -- on a laptop it takes 0.04~s to determine the user's spatial and time coordinates with a relative accuracy of $10^{-28}-10^{-26}$ and $10^{-32}-10^{-30}$, respectively.
\end{abstract}
%
\submitto{\CQG}
%
%
%
%
%
%
%
\section{Introduction}
\label{sec:introduction}
Current Global Navigation Satellites Systems (GNSS), such as the Global Positioning System and the European Galileo system, are based on Newtonian concept of absolute space and time. The signals from four satellites are needed for a receiver to determine its position and time via the time difference between the emission and the reception of the signal. This concept of absolute space and time would work ideally if satellites and the receiver were at rest in an inertial reference frame. It is also a good approximation for a slowly moving receiver (with velocity $v\ll c$) in a very weak gravitational field.

However, at the level of precision needed by current GNSS, space and time around Earth can not be considered as absolute and the effects of inertial reference frames and curvature of the space-time in the vicinity of Earth have to be taken into account. Relativistic effects are far from being negligible \cite{Ashby2003,Pascual-Sanchez2007}; the most important ones are the gravitational frequency shift between clocks and the Doppler shift of the second order -- for Galileo GNSS, they amount to around $38.5\ \mathrm{\mu s}$, corresponding to 12 km error in satellite position after one day of integration. Since this is much higher than the required precision, it is obvious that relativistic effects have to be included in the description of the GNSS.

There are two ways of including relativity in the description of GNSS: one way is to keep the Newtonian concept of absolute time and space, and add a number of relativistic corrections to the level of the desired accuracy. An alternative and more consistent approach is to abandon the concept of absolute space and time and describe a GNSS directly in general relativity, i.e. to define a \emph{Relativistic Positioning System} (RPS) with the so-called emission coordinates in the following way \cite{Coll1991, Rovelli2002, Blagojevic2002, Coll2003, Tarantola2009}. 

\begin{figure}
\centering
\includegraphics[height=0.4\textwidth]{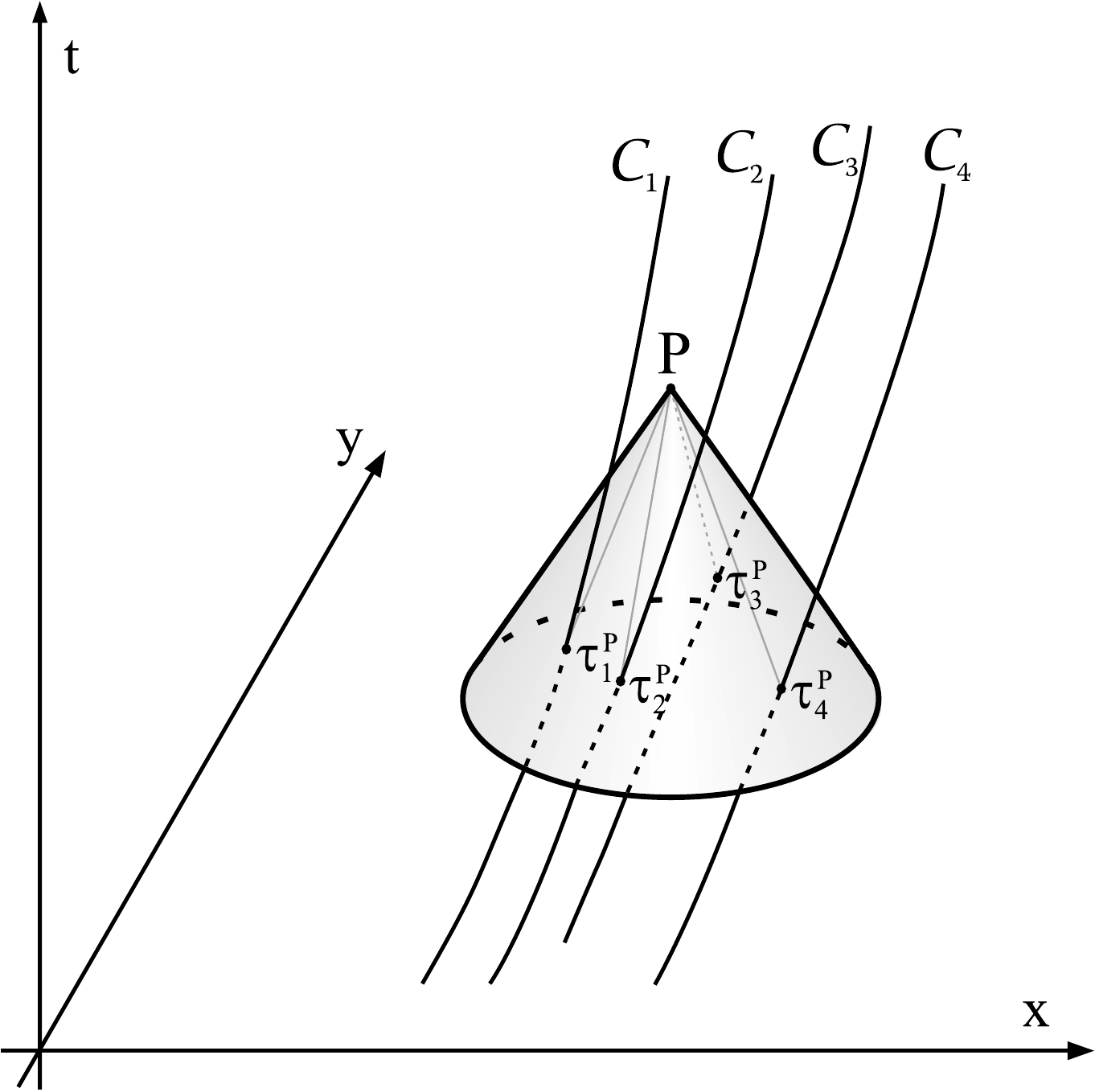}
\hspace{1cm}
\includegraphics[height=0.4\textwidth]{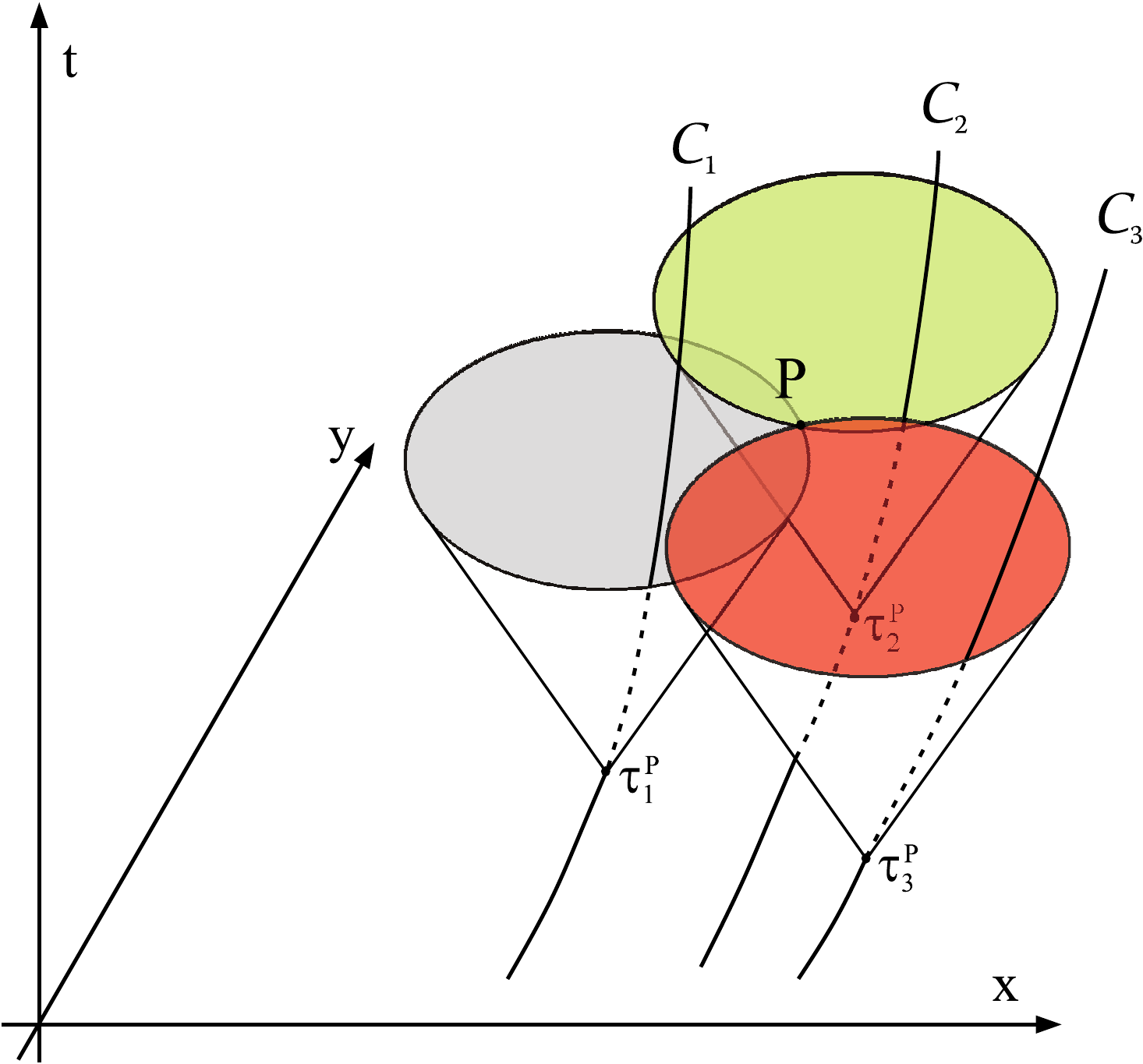}
\caption{Left: $\cal{C}$$_i$ is the worldline of a particle $i$ parametrized by its proper time $\tau_i$; its origin $O$ is in $\tau_i=0$.
Past null cone of the event $P$ crosses the worldline $\cal{C}$$_i$ of the particle $i$ at its proper time $\tau_i^P$. Right: Each worldine $\cal{C}$$_i$ defines a one-parameter family of future null cones.}
\label{fig:em_coord}
\end{figure}

Let us have four particles $i = 1, 2, 3, 4$. Their worldlines $\cal{C}$$_i$ are parametrized by their proper times $\tau_i$. Let $P$ be an arbitrary event. The \emph{past} null cone of $P$ crosses each of the four worldlines $\cal{C}$$_i$ in $\tau_i^P$ (see \reff{fig:em_coord}, left). Having four particles with four wordlines $\cal{C}$$_1$, $\cal{C}$$_2$, $\cal{C}$$_3$, $\cal{C}$$_4$, the past cone of $P$ crosses them at $\tau_1^P, \tau_2^P, \tau_3^P, \tau_4^P$. Then $(\tau_1^P, \tau_2^P, \tau_3^P, \tau_4^P)$ are the emission coordinates defining the event $P$.

We can see this also in a different way. The worldline $\cal{C}$$_i$ of the particle $i$ defines a one-parameter family of {\em future} null cones, which can be parametrized by proper time $\tau_i$ (see \reff{fig:em_coord}, right). The intersection of four future null cones from four worldlines $\cal{C}$$_i$ at $\tau_i^P$ defines an event with coordinates $\tau_1^P, \tau_2^P, \tau_3^P, \tau_4^P$. Position of event $P$ is therefore defined in this particular coordinate system.

Now we consider that the four particles are four satellites broadcasting their proper time. A user of an RPS receives at a given moment (event $P$) four signals from four different satellites and is able to determine the proper time $\tau_i^P$ of each satellite at the moment of emission of the signal. These four proper times $(\tau_1^P, \tau_2^P, \tau_3^P, \tau_4^P)$ therefore constitute its \emph{emission coordinates}. By receiving them at subsequent times, the receiver knows its trajectory in the emission coordinates.

An RPS with emission coordinates in Schwarzschild space-time was modelled in \cite{vCadevz2010}, where it was demonstrated that relativistic approach enables the construction of a very accurate GNSS: the dynamics of the satellites, the calculation of the user's emission coordinates, and the transformation from the emission coordinates to the more customary Schwarzschild coordinates were treated in the formalism of general relativity, where all the algorithms provided very high accuracy.

In this paper we model an RPS in a more realistic space-time in vicinity of Earth, which includes all relevant gravitational perturbations: Earth multipoles up to $6^{\mathrm{th}}$, the Moon, the Sun, Jupiter, Venus, solid and ocean tides, and Kerr effect. The emphasis of the paper is on the relativistic approach to global positioning, and the precision and novelty of the Earth's model is not our primary concern as it does not affect the concept of RPS. We used data available for coordinate systems and solid tides in ITRS \cite{Petit2010}, for ocean tides in \cite{Biancale2012} and for average Earth's multipole moments in EGM96 \cite{Lemoine1998}. We want to stress that we did not include any non-gravitational perturbations in the model (e.g. problems associated with geophysics, signal propagation, radiation pressure, clock noise...), even though some of them may induce larger effects than some gravitational perturbations, because it is not well known how to include them in the formalism of general relativity. The treatment of non-gravitational perturbations within RPS is certainly necessary and deserves further study, however it is beyond the scope of this paper.

In order to construct an RPS, we calculate satellite dynamics in the perturbed space-time. For this purpose, we derive perturbed metric in \refsec{sec:metric}. In \refsec{sec:evolution_of_orbits}, we solve the geodesic equation and present some results on time evolution of orbital parameters of satellites. In \refsec{sec:RPS}, we model a satellite positioning system by simulating satellites' orbits, emission of their signals, and by calculating the emission coordinates of the observer, which can then be transformed into a more customary set of spherical coordinates $(t,r,\theta,\phi)$.

\section{Perturbed metric}
\label{sec:metric}
We write the metric of the space-time around the Earth $g_{\mu\nu}$ as a sum of spherically symmetric and time independent background given by the Schwarzschild metric $g^{(0)}_{\mu \nu}$ and the metric perturbations $h_{\mu \nu}$:
\beq
  g_{\mu\nu} = g_{\mu\nu}^{(0)} + h_{\mu\nu}\>.
  \label{eq:metric_pert}
\eeq
The gravitational perturbations are several orders of magnitude smaller than the Earth's gravitational $GM$ term: $h_{\mu \nu} \ll g^{(0)}_{\mu \nu}$. We are interested in the space-time outside Earth, therefore the perturbative metric $h_{\mu \nu}$ must satisfy the Einstein equation for the vacuum:
\beq
  \delta R_{\mu\nu} = 0 \>,
\eeq
where $\delta R_{\mu\nu}$ is the Ricci tensor due to metric perturbation. The linearized Einstein equations, where we neglect the second order metric perturbation contributions $O(h^2)$, are equal to
\beq
h_{\alpha\phantom{\alpha};\mu\nu}^{\phantom{\alpha}\alpha} - h_{\mu\phantom{\alpha};\nu\alpha}^{\phantom{\mu}\alpha} -
h_{\nu\phantom{\alpha};\mu\alpha}^{\phantom{\nu}\alpha} + h_{\mu\nu;\phantom{\alpha}\alpha}^{\phantom{\mu\nu;}\alpha} 
+g_{\mu\nu}^{(0)}(h_{\alpha\phantom{\alpha};\phantom{\lambda}\lambda}^{\phantom{\alpha}\lambda\phantom{;}\alpha} - h_{\lambda\phantom{\lambda};\alpha}^{\phantom{\lambda}\lambda\phantom{;\alpha}\alpha} )
 = 0\>,
\label{equationA}
\eeq
where a semi-colon (;) denotes covariant derivative, calculated with respect to the unperturbed metric $g_{\mu\nu}^{(0)}$. Due to omission of the terms of the order of $O(h^2)$ in the metric, it is clear that the solutions of these equations can approximate the metric only to the first order in the metric perturbation. Although a general formalism for solving \refe{equationA} and thus including perturbations in Schwarzschild space-time already exists \cite{2005PhRvD..72l4016F,2005PhRvD..71j4003M}, we expand the perturbative metric into multipoles using Regge-Wheeler-Zerilli formalism \cite{Regge1957, Zerilli1970} to find solutions appropriate for relativistic positioning system. (Details are given in \ref{app:rwz}.)

In this formalism, the metric perturbation $h_{\mu\nu}$ is expressed as a series of expansion terms $(h_{\mu\nu}^{nm})^{(\rm o)}$ and $(h_{\mu\nu}^{nm})^{(\rm e)}$, corresponding to the odd-parity and even-parity normal modes, respectively. Based on the locations of the sources of perturbations, these modes can be grouped into two terms:
\beq
  h_{\mu\nu} =  h_{\mu\nu}^\oplus + h_{\mu\nu}^\ominus\>.
\eeq
The first term, $h_{\mu\nu}^\oplus$, represents the asymptotically flat metric perturbation associated to the Earth's time dependent (exterior) spherical multipole momenta and the frame-dragging effect of the Earth via the Kerr contribution, whereas the second term, $h_{\mu\nu}^\ominus$, is not asymptotically flat and describes the metric perturbation due to celestial bodies with their frame-dragging effect neglected. (Both terms are discussed in \ref{sec:timedep}.)

To simplify expressions, we introduce the normalized complex spherical multipoles $\overline{M}_{nm}^{\oplus}$, $\overline{M}_{nm}^{\ominus}$ defined in \ref{app:multipoles} as
\begin{align}
  \overline{M}_{nm}^{\oplus}  &:= \frac{2}{c^2}  M_{nm}^{\oplus}\\
  \overline{M}_{nm}^{\ominus}  &:= \frac{2}{c^2}  M_{nm}^{\ominus}\ .
\end{align}
We note that multipole coefficients $\overline{M}_{nm}$ are expansion coefficients of the Newtonian potential, and are only the leading order approximations of exact relativistic coefficients in the limit $c\rightarrow \infty$ (see e.g. \cite{Quevedo1990, Backdahl2005}).

Taking into account the results from \ref{app:rwz} and \ref{app:multipoles}, the metric perturbations are expressed in the following compact form.

Metric perturbation due to the Earth's multipoles and rotation can be written as
\beq
  \begin{split}
  &[h_{\mu\nu}^\oplus] = \sum_{nm} \overline{M}_{nm}^\oplus Y_n^m\cdot
   {\rm diag} 
   \left (
   \frac{P_n^{(0)}}{r^{n+1}},
   \frac{P_n^{(0)}}{r^{n-1}(r-r_{\rm s})^2},
   \frac{P_n^{(1)}}{r^{n-1}},
   \frac{P_n^{(1)}\sin^2\theta}{r^{n-1}} 
   \right)\\
    &+\sum_{nm}{\overline{M}_{nm}^\oplus}_{,T}
    Y_n^m \frac{P_n^{(3)}}{r^{n-1}(r-r_{\rm s})}
   (\delta_{\mu,1}\delta_{\nu,0}+ \delta_{\nu,1}\delta_{\mu,0})\\
  &- a \frac{r_{\rm s}}{r}\sin^2\theta  
  (\delta_{\mu,3}\delta_{\nu,0}+ \delta_{\nu,3}\delta_{\mu,0})\ ,
  \end{split}
  \label{eq:h_earth1}
\eeq
where the Earth's multipoles $\overline{M}_{nm}^\oplus$ are functions of time and include rotation and tides, and ${\overline{M}_{nm}^\oplus}_{,T}$ are their time derivatives. Therefore, the first two terms in (\ref{eq:h_earth1}) describe the metric perturbation due to oscillating multipoles and tides, while the third describes the Kerr frame-dragging effect.

Metric perturbations due to celestial bodies are
\beq
  \begin{split}
  &[h_{\mu\nu}^\ominus] = \sum_{nm} \overline{M}_{nm}^\ominus Y_n^m\cdot
    {\rm diag} \left (
   r^n R_n^{(0)},
   \frac{r^{n+2}R_n^{(0)}}{(r-r_{\rm s})^2},
   r^{n+2}R_n^{(1)},
   r^{n+2}R_n^{(1)}\sin^2\theta
   \right)\\
    &+\sum_{nm} {\overline{M}_{nm}^\ominus}_{,T}
   Y_n^m \frac{r^{n+2} R^{(3)}}{r-r_{\rm s}} 
   (\delta_{\mu,1}\delta_{\nu,0}+ \delta_{\nu,1}\delta_{\mu,0})\>,
   \end{split}
  \label{eq:h_celestial1}
\eeq
where $\overline{M}_{nm}^\ominus$ are summed multipoles of celestial bodies.

The first order approximations of the metric perturbations given by (\ref{eq:h_earth1}) and (\ref{eq:h_celestial1}) are fully determined by multipole momenta $\overline{M}_{nm}^\oplus$, $\overline{M}_{nm}^\ominus$, Kerr parameter $a$, and functions $P_n^{(i)}$ and $R_n^{(i)}$. 

\section{Dynamics of satellites}
\label{sec:evolution_of_orbits}
%
%
%
%
%
%
%
To calculate a perturbed satellite orbit, we integrate the geodesic equation
\beq
  \frac{\dd^2 x^\mu }{\dd \tau^2} + 
  \Gamma^{\mu}_{\phantom{\mu}\alpha\beta}
  \frac{\dd x^\alpha}{\dd \tau}\frac{\dd x^\beta }{\dd \tau} = 0 \>,
  \label{eq:geodesic}
\eq
where $\Gamma^c_{\phantom{c}ab}=\frac{1}{2} g^{cd}(g_{da,b}+g_{db,a}-g_{ab,d})$ are the Christoffel symbols of the second kind and $\tau$ is the proper time. The Christoffel symbols are calculated using the perturbed Schwarzschild metric $g_{\mu\nu}$ as given in \refsec{sec:metric}, whereby we do the following considerations:

The metric perturbation associated to the Kerr effect is already well known and provided in \ref{sec:odd_timeindep}. We verified that only the effect of rotation of the Earth's monopole is large enough for required accuracy (see also \cite{Hartle1967,Hartle1967a}), thus, we neglected the effect of rotating higher multipoles. The contributions of the Moon, the Sun, Venus, and Jupiter to the metric perturbation are described by a multipole expansion, where the multipole momenta are calculated from the positions of these objects. Perturbations due to the Earth multipoles are treated as shown in the previous section, while the tidal effects of the Sun and the Moon on the Earth's crust and oceans are modelled with time changing Earth multipoles, described in \ref{app:multipoles}. 

In both cases, the positions of celestial bodies were obtained from the ephemerides \cite{Horizons2013}. They were sampled with a time step of 1 hour, from 1 January 2012 to 31 December 2012. However, in the numerical integration of \refe{eq:geodesic} a much finer time sampling is required, therefore, we use an interpolation of multipoles as a function of time for intermediate times. 

To make sure that even the weakest contributions are not lost in numerical noise, the calculations had to be very accurate, while on the other hand, we wanted them to be as time efficient as possible. To meet both criteria, we used the 8th order B-spline interpolation and 8th or 10th order explicit Runge-Kutta integration scheme, implemented in 128-bit floating point arithmetics.

Initial values of orbital parameters were: time of the first apoapsis passage $t_a = 7\ \mathrm{h}$, argument of the apoapsis $\omega=0^{\circ}$, longitude of the ascending node $\Omega = 0^{\circ}$, semi-major axis $a=29600\ \mathrm{km}$, eccentricity $\ecc = 0.007$, inclination $\incl=56^{\circ}$ (the values of $a$ and $\incl$ are in agreement with the Galileo GNSS). We integrated \refe{eq:geodesic} from this initial position, and for every new position of the satellite we also calculated new orbital parameters to obtain their evolution. We performed orbit calculations for each perturbation separately and for the sum of all perturbations.
%
%
%
%
\begin{figure}
\centering
	\includegraphics[width=0.42\textwidth]{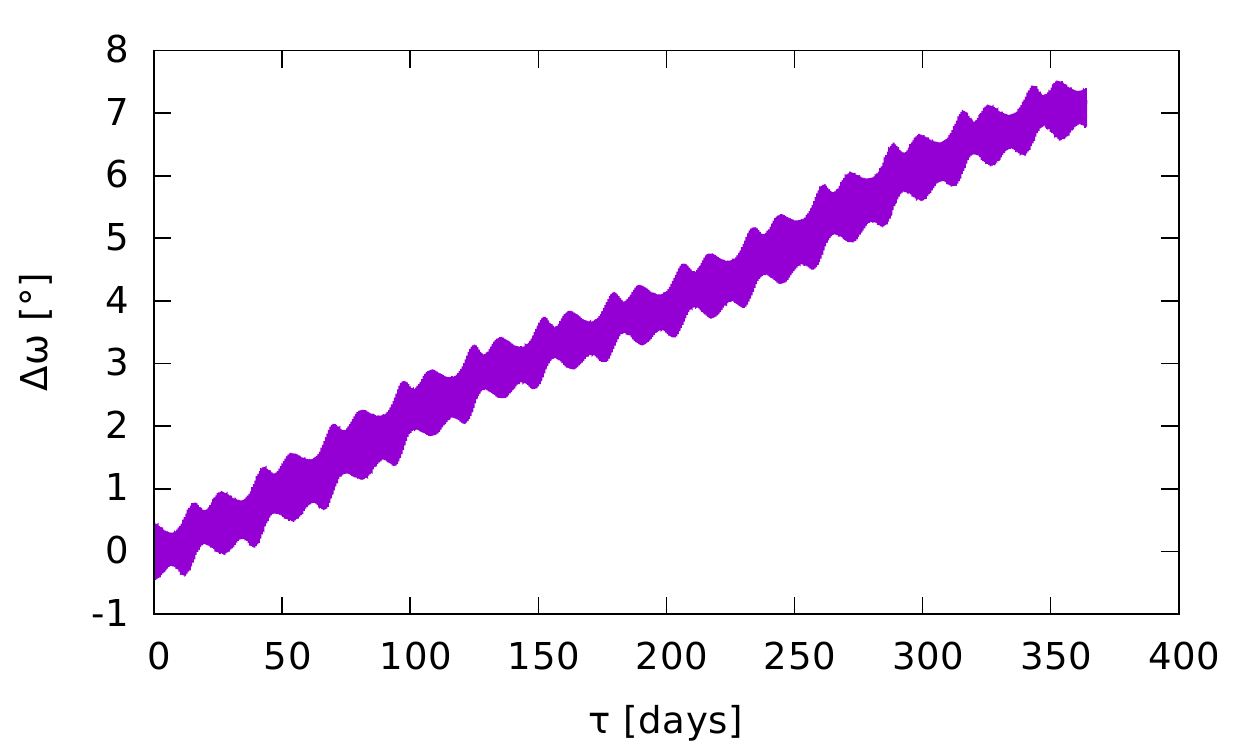}
	\includegraphics[width=0.42\textwidth]{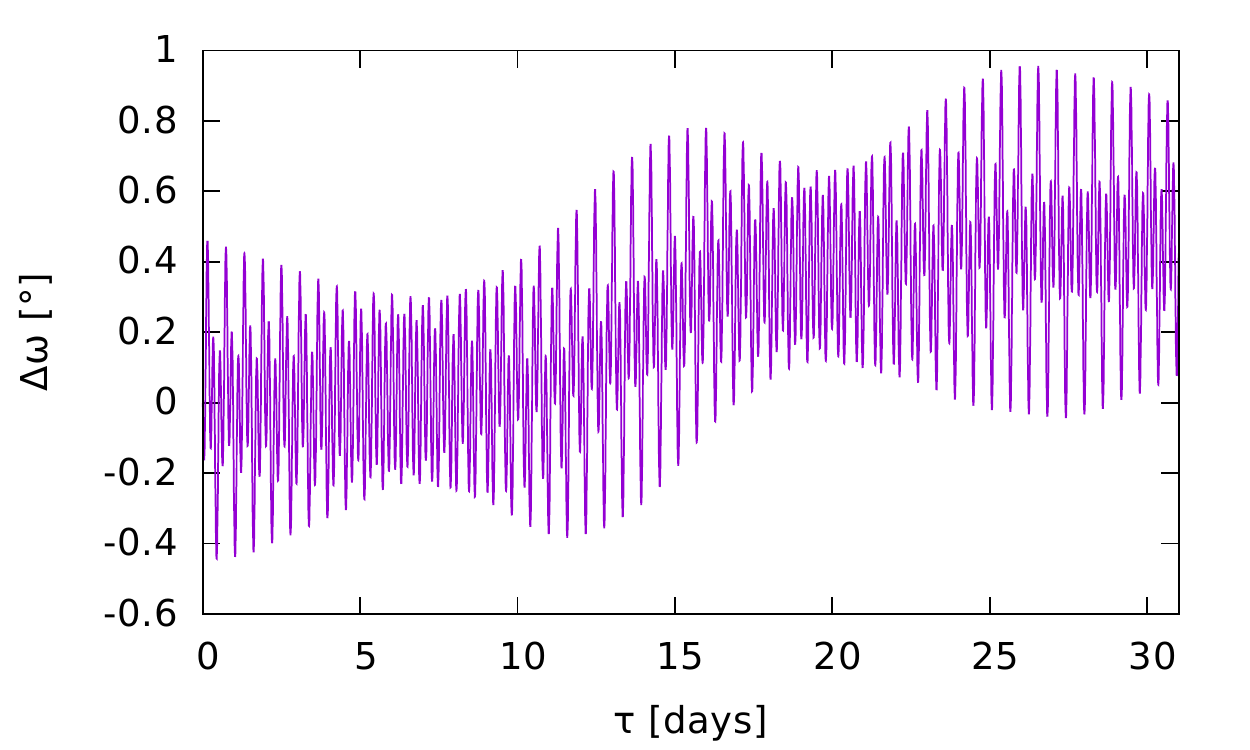}\\
	\includegraphics[width=0.42\textwidth]{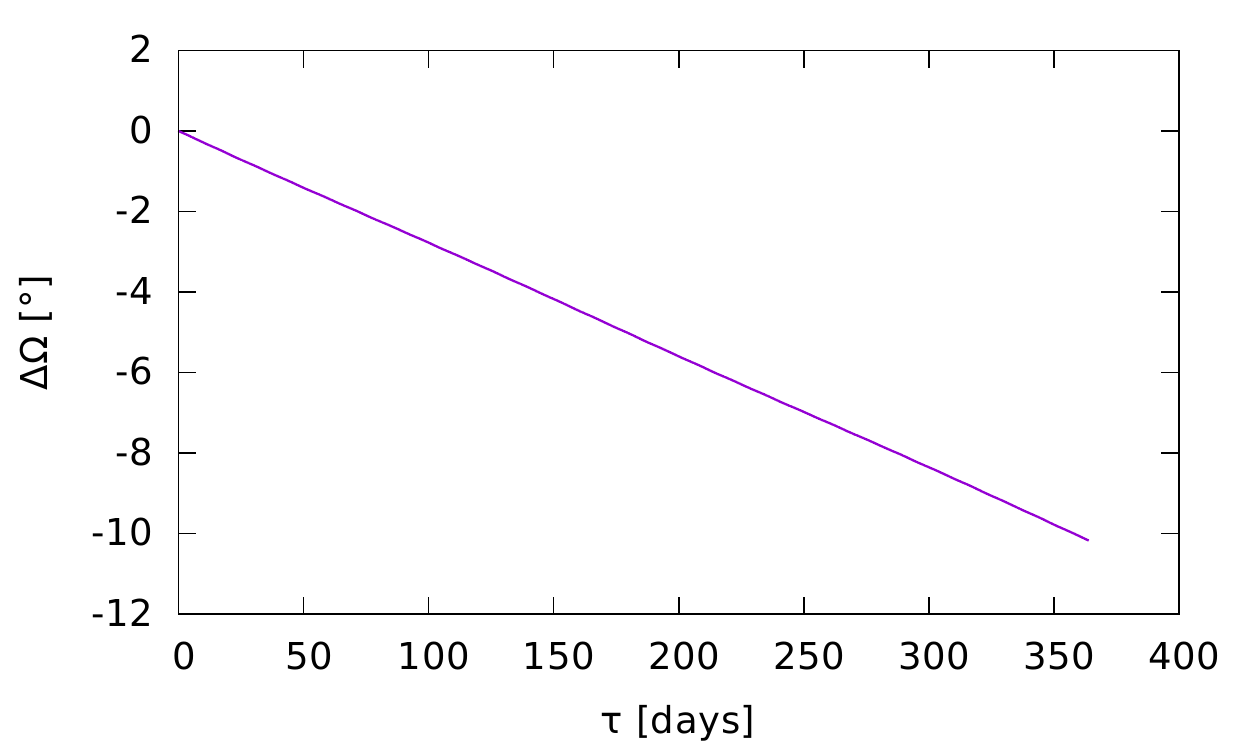}
	\includegraphics[width=0.42\textwidth]{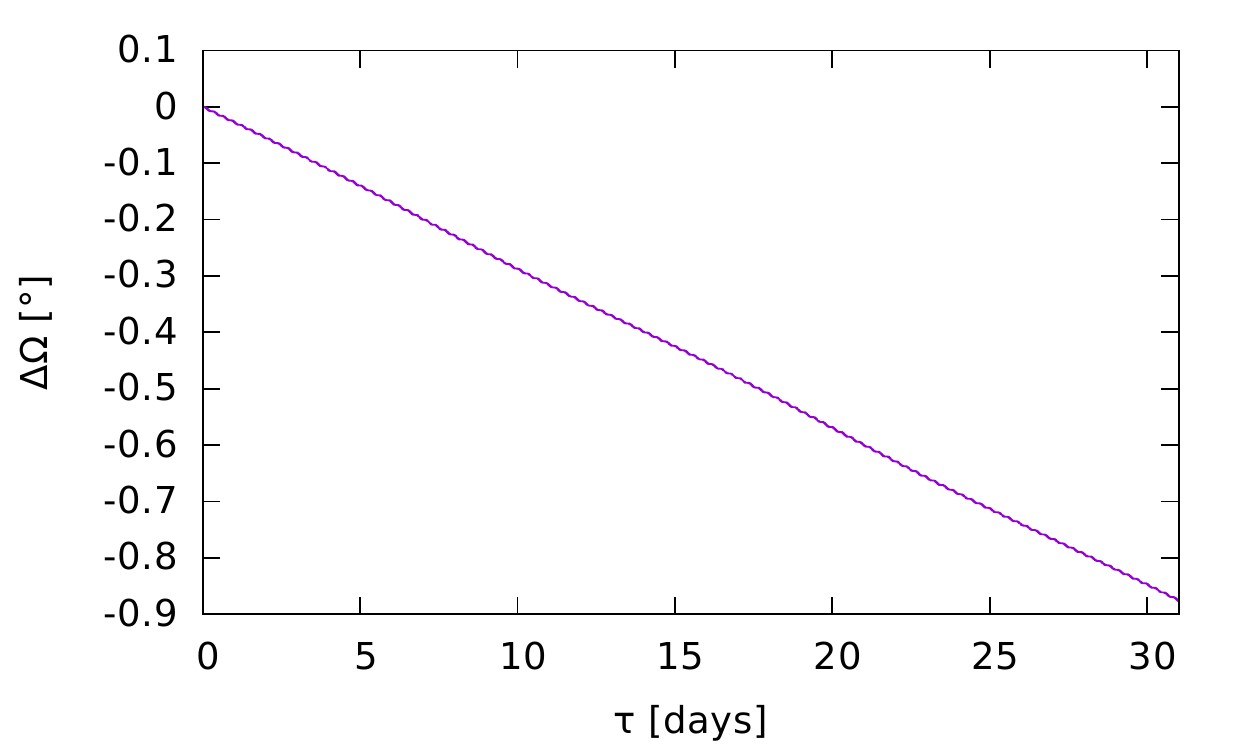}\\
	\includegraphics[width=0.42\textwidth]{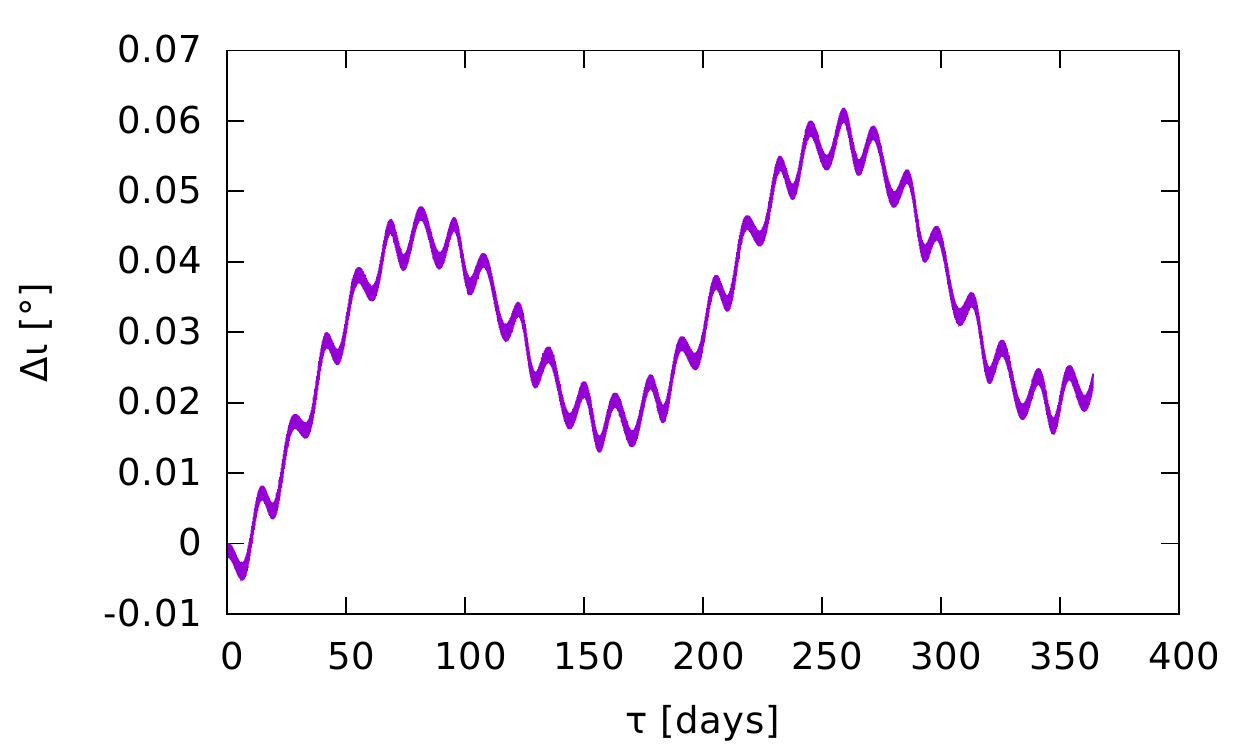}
	\includegraphics[width=0.42\textwidth]{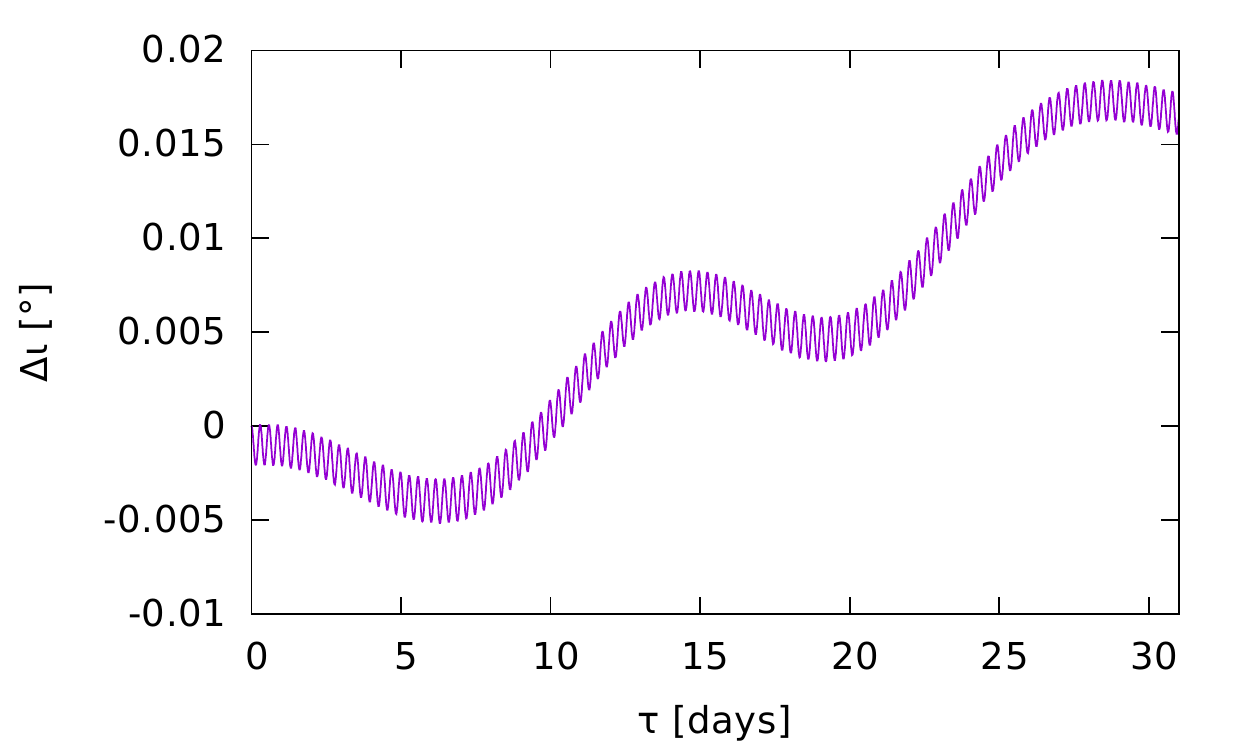}\\
	\includegraphics[width=0.42\textwidth]{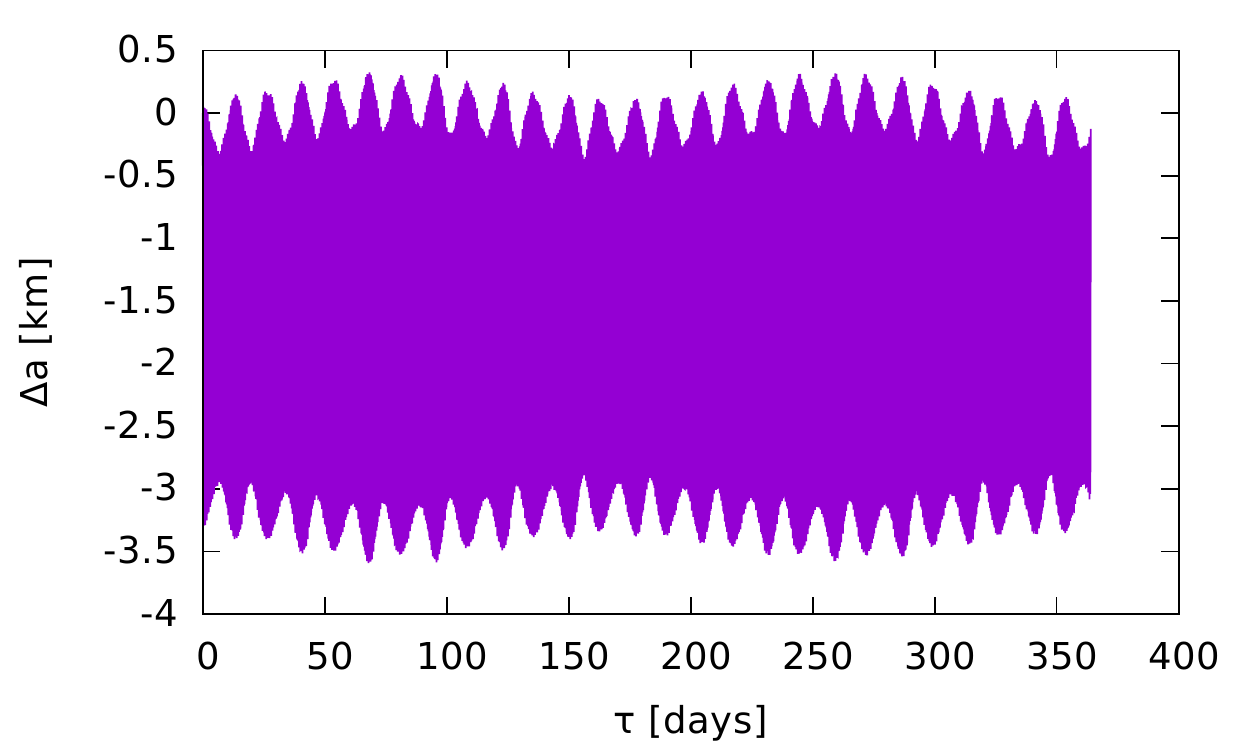}
	\includegraphics[width=0.42\textwidth]{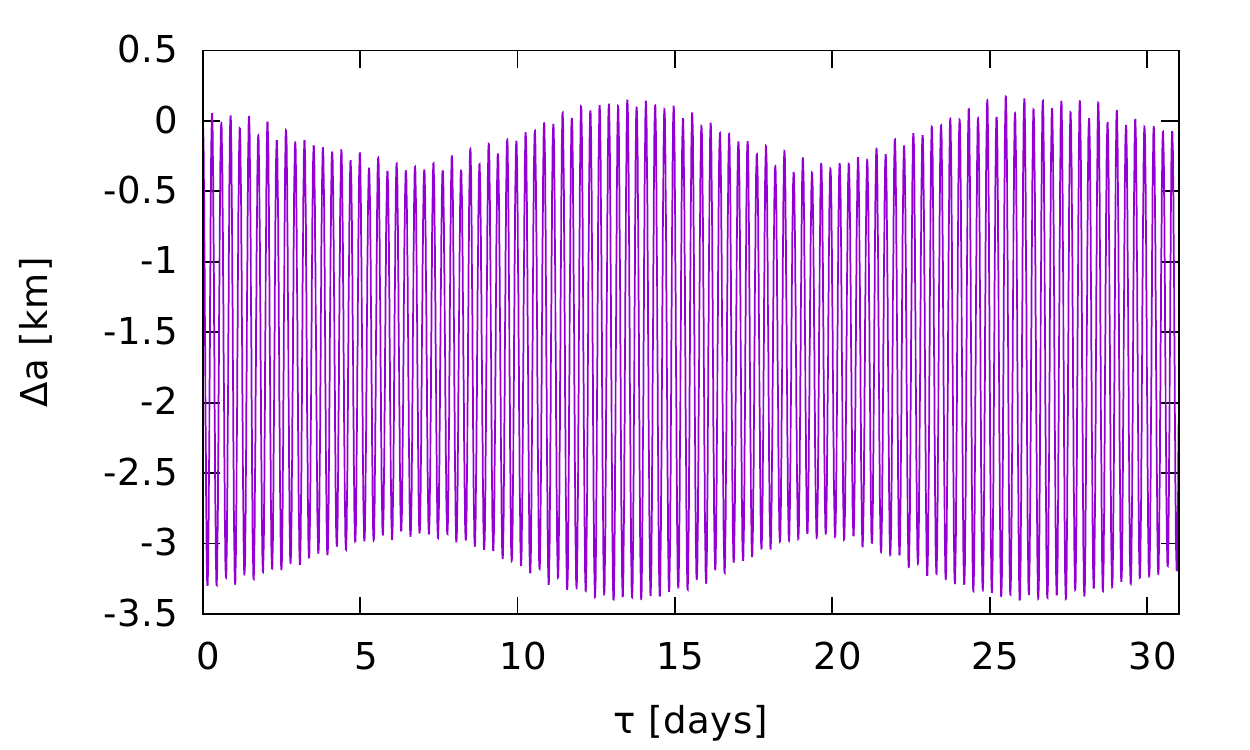}\\
	\includegraphics[width=0.42\textwidth]{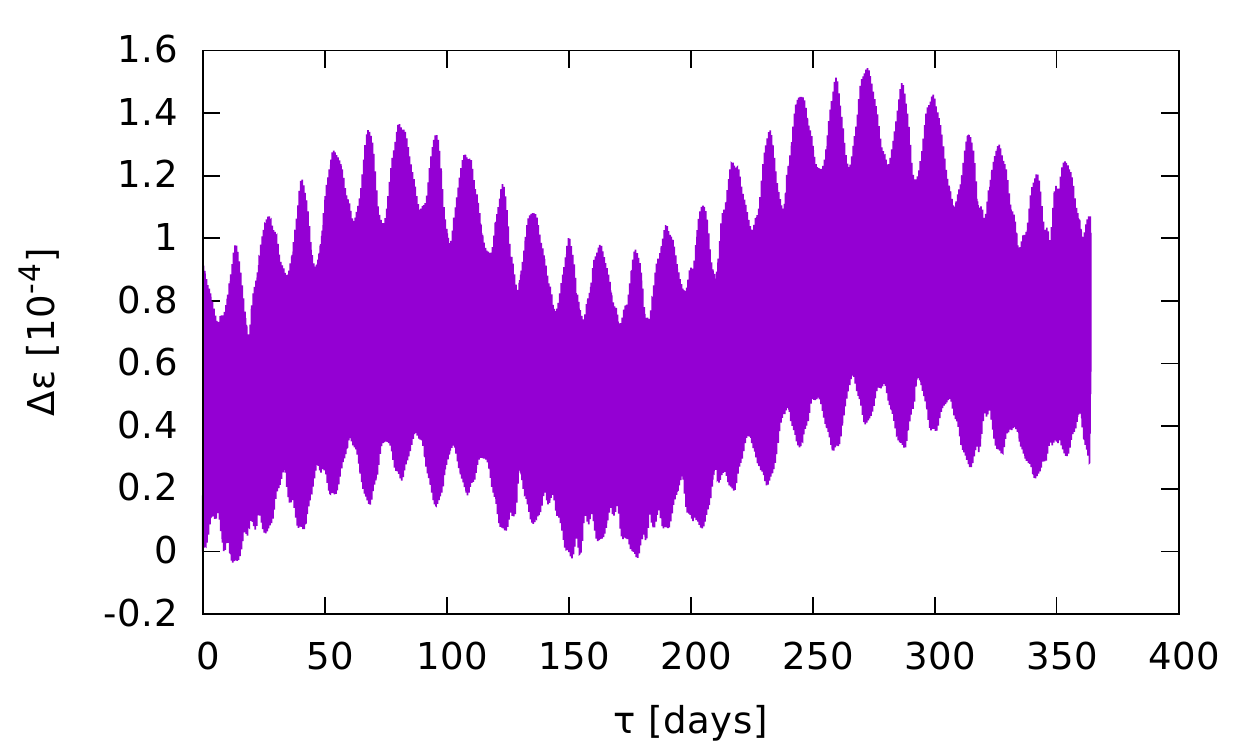}
	\includegraphics[width=0.42\textwidth]{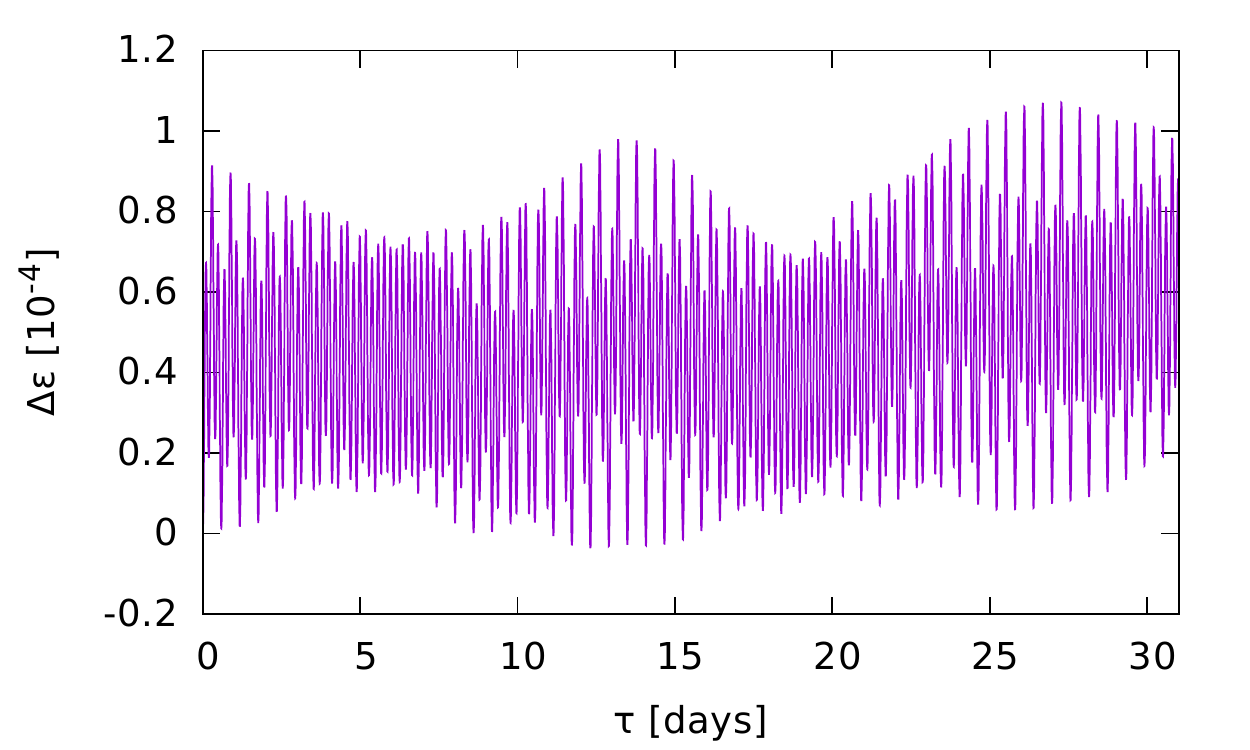}
\caption{Evolution of orbital parameters due to sum of all perturbations (Earth multipoles, Earth solid and ocean tide, the Moon, the Sun, Venus, Jupiter, and Kerr effect). The graphs on the left show the long-term changes of the orbital parameters in one year, while the graphs on the left show the short time-scale changes within the first 7 days. The time on $x$-axis counts days from 1 January 2012 at 7:00 a.m.}
\label{fig:AllParameters}
\end{figure}
\subsubsection*{Effects of perturbations on orbital parameters}
Our modelling shows that the contributions of metric perturbations can be separated into an oscillating and a secular term (\reff{fig:AllParameters}). As expected, the effects due to Earth multipoles and the Moon produce the largest variations in orbital parameters. The amplitudes of the oscillations for Earth multipoles are: $\Delta \omega \approx 0.4^\circ$, $\Delta \Omega \approx 18''$, $\Delta \incl \approx4''$, $\Delta a \approx 1.5\ \mathrm{km}$, and $\Delta \ecc \approx 4\times 10^{-5}$, while for the Moon they are: $\Delta \omega \approx 0.8^\circ$, $\Delta \Omega \approx 1''$, $\Delta \incl \approx 0.7''$, $\Delta a \approx 300\ \mathrm{m}$, and $\Delta \ecc \approx 10^{-5}$ (see Table \ref{tab:paramp}). The secular changes due to Earth multipoles are: $\A{\omega}\approx 5^\circ/\mathrm{yr}$, $\A{\Omega} \approx -9.5^\circ/\mathrm{yr}$,  $\A{\incl} = 0$, $\A{a} = 0$, and $\A{\ecc} \approx 2\times 10^{-5}/\mathrm{yr}$, and for the Moon they are: $\A{\omega}\approx 1.75^\circ/\mathrm{yr}$, $\A{\Omega} \approx -0.55^\circ/\mathrm{yr}$,  $\A{\incl}\approx 3'/\mathrm{yr}$, $\A{a} = 0$, and $\A{\ecc} \approx 2\times 10^{-5}/\mathrm{yr}$ (see Table \ref{tab:pardot}).

The smallest variations in parameters come from the Kerr effect, with the amplitudes of the oscillations of $\Delta \omega \approx 3.6\times 10^{-2}$ mas and $\Delta \Omega \approx 10^{-3}\ \mathrm{mas}$. In this case, the secular contributions are $\A{\omega} \approx -4\ \mathrm{mas/yr}$, and $\A{\Omega}\approx 2.5\ \mathrm{mas/yr}$, while the remaining parameters have no secular changes.

Results show that the effects of perturbations on the orbital parameters are very small, therefore we can treat perturbed orbits as approximately planar, i.e. the orientation of the orbital plane slowly changes with time.

\begin{table}
\caption{\label{tab:paramp}The amplitudes of oscillations of orbital parameters due to perturbations.}
\begin{indented}
\lineup
\item[]\begin{tabular}{@{}lccccc}
  \br
    perturbation & $\Delta\omega$ & $\Delta \Omega$ & $\Delta\incl$ & $\Delta a$ & $\Delta \ecc$\\
  \mr
 Earth multipoles & $0.4^{\circ}$ & $18''$ & $4''$ & $1.5\ \mathrm{km}$ & $4\times 10^{-5}$\\
  solid tide & $0.05''$ & $2\times 10^{-4}\phantom{ }''$ & $10^{-4}\phantom{ }''$ & $5\ \mathrm{cm}$ & $4\times 10^{-9}$\\
  ocean tide & $10^{-3}\phantom{ }''$ & $3\times 10^{-6}\phantom{ }''$ & $3\times 10^{-6}\phantom{ }''$ & $0.6\ \mathrm{mm}$ & $3\times 10^{-11}$\\
  Moon & $0.8^{\circ}$ & $1''$ & $0.7''$ & $300\ \mathrm{m}$ & $10^{-5}$ \\
  Sun & $2'$ & $0.7''$ & $0.2''$ & $100\ \mathrm{m}$ & $5\times 10^{-6}$\\
  Venus & $5\times 10^{-5}\phantom{ }''$ & $4\times 10^{-7}\phantom{ }''$ & $4\times 10^{-7}\phantom{ }''$ & $0.2\ \mathrm{mm}$ & $10^{-11}$\\
  Jupiter & $2\times 10^{-3}\phantom{ }''$ & $2\times 10^{-6}\phantom{ }''$ & $3\times 10^{-6}\phantom{ }''$ & $1\ \mathrm{mm}$ & $5\times 10^{-11}$\\
  Kerr & $3.6\times 10^{-5}\phantom{ }''$ & $10^{-6}\phantom{ }''$ & $3\times 10^{-6}\phantom{ }''$ & $5\times 10^{-14}\ \mathrm{m}$ &  $2\times 10^{-12}$\\
 \br
  \end{tabular}%
\end{indented}
\end{table}
\begin{table}
\caption{\label{tab:pardot}The secular contribution of perturbations to evolution of orbital parameters. The values are per year.}
\begin{indented}
\lineup
\item[]\begin{tabular}{@{}lccccc}
  \br
  perturbation & $\A{\omega}$ & $\A{\Omega}$ & $\A{\incl}$ & $\A{a}$ & $\A{\ecc}$\\
  \mr 
 Earth multipoles & $5^{\circ}$ & $-9.5^{\circ}$ & 0 & 0 & $2\times 10^{-5}$\\
  solid tide & $0.2''$ & $-0.3''$ & 0 & 0 & 0\\
  ocean tide & 0 & 0 & 0 & 0 & 0\\
  Moon & $1.75^{\circ}$ & $-0.55^{\circ}$ & $3'$ & 0 & $2\times 10^{-5}$\\
  Sun & $0.7^{\circ}$ & $-0.25^{\circ}$ & 0 & 0 & 0\\
  Venus & $0.05''$ & $0.025''$ & $6\times 10^{-3}\phantom{ }''$ & 0 & $1.6\times 10^{-9}$\\
  Jupiter & $0.007''$ & $-0.01''$ & $-0.005''$ & 0 & $1.2\times 10^{-9}$\\
  Kerr & $-0.004''$ & $2.5\times 10^{-3}\phantom{ }''$ & 0 & 0 & 0\\
 \br
  \end{tabular}%
\end{indented}
\end{table}

\begin{figure}
\centering
  \includegraphics[width=0.45\linewidth]{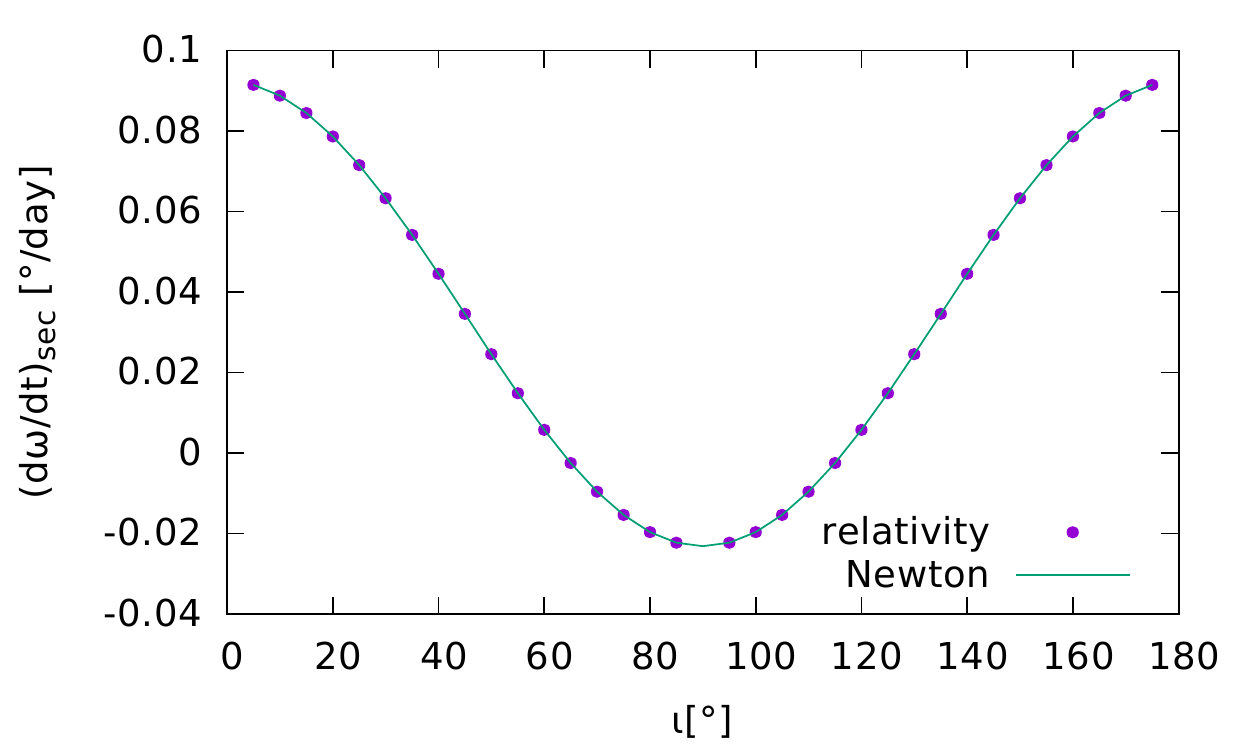}
	\includegraphics[width=0.45\linewidth]{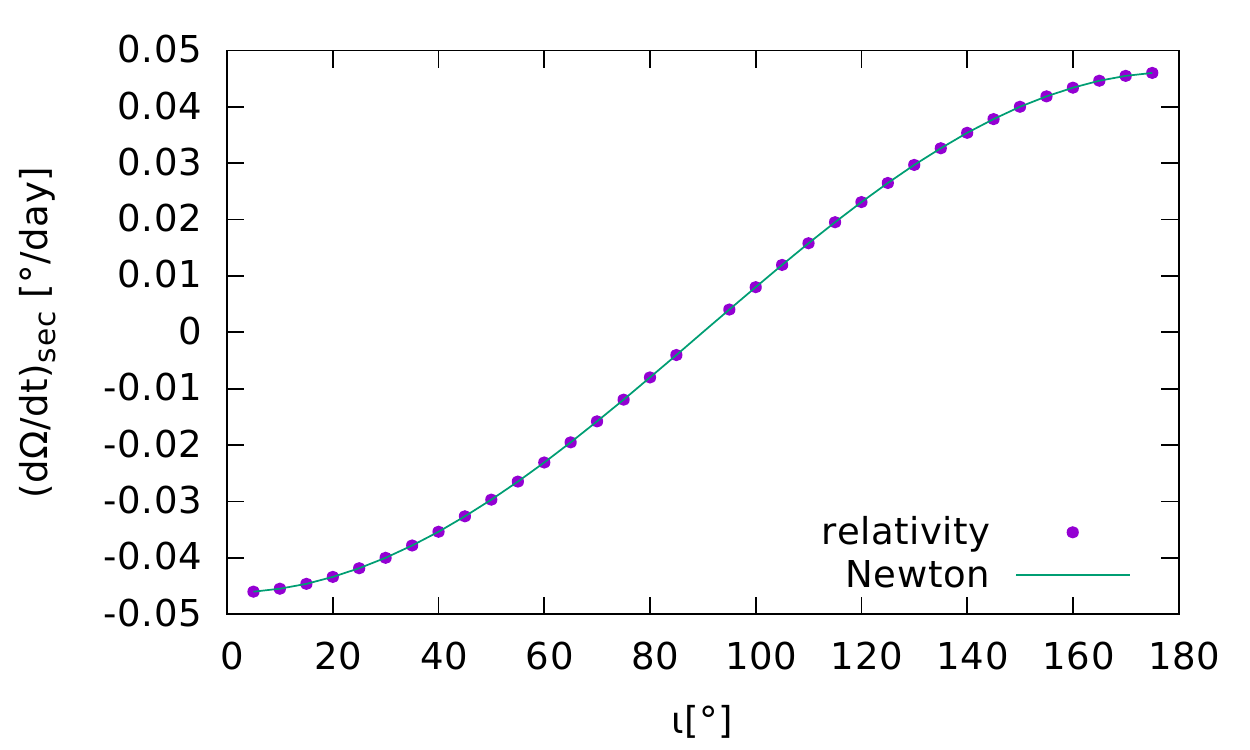}
\caption{Comparison of relativistic numerical calculation and Newtonian analytical estimate of secular drift in longitude of the ascending node $\Omega$ (left) and argument of periapsis $\omega$ (right) due to Earth quadrupole. $\incl$ is the starting satellite's inclination.}
\label{fig:scan_inclination}
\end{figure}

We compared our results for $\omega$ and $\Omega$ to Newtonian secular evolution of these parameters, which for Earth quadrupole have the following dependence on inclination \cite{Blitzer1970}
\beq
  \B{\omega} \propto 4-5 \sin^2 \incl \hspace{0.5cm} \mathrm{and}\hspace{0.5cm} 
  \B{\Omega} \propto \cos \incl
\label{eq:newton}
\eq
for low eccentricity orbits. To obtain the secular drift, we simulated the satellite dynamics for various initial inclinations and performed a least-square fitting of a line to their time evolution. The directional coefficient of the line $k_{\omega}$ and $k_{\Omega}$ should be an empirical approximation of the secular drift
\begin{equation}
  k_{\omega} \approx \B{\omega} \hspace{0.5cm} \mathrm{and}\hspace{0.5cm}
  k_{\Omega} \approx \B{\Omega}\ .
\end{equation}
As shown in \reff{fig:scan_inclination}, our results agree with analytical approximations \refe{eq:newton}.
\subsubsection*{Effects of perturbations on the position and time of the satellite}
\begin{figure}
\centering
	\includegraphics[width=0.45\linewidth]{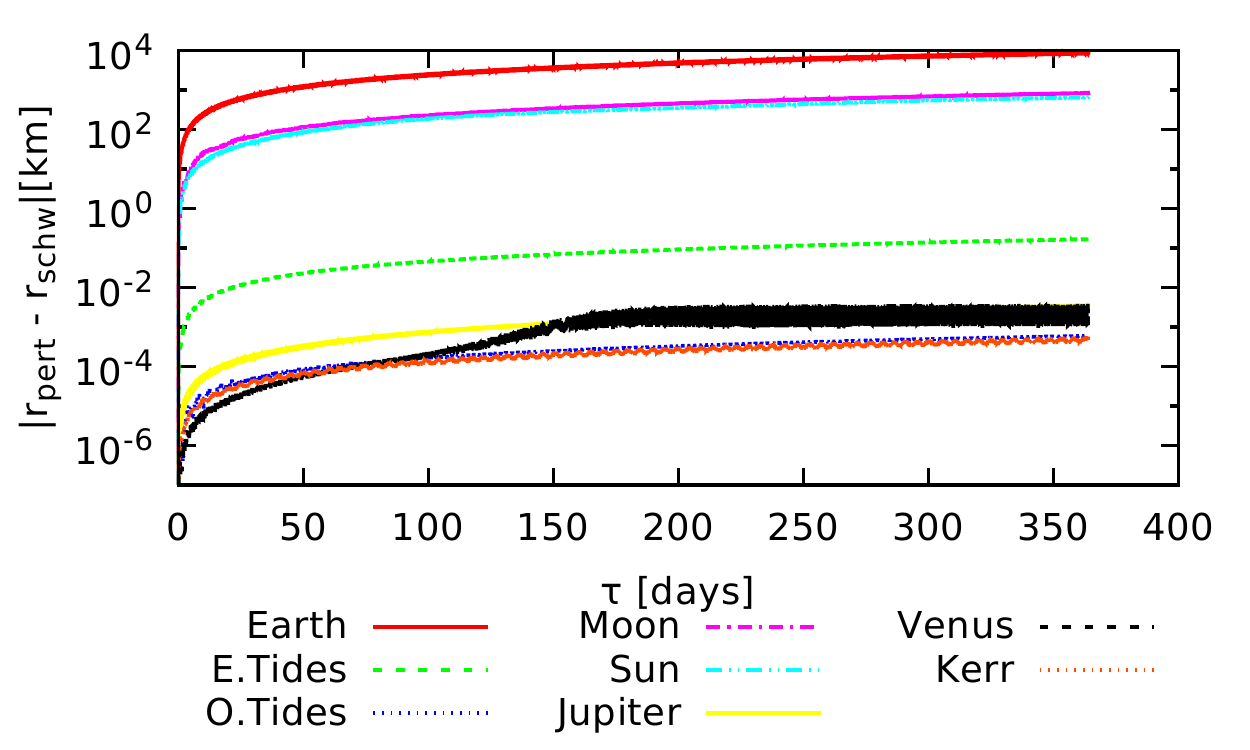}
	\includegraphics[width=0.45\linewidth]{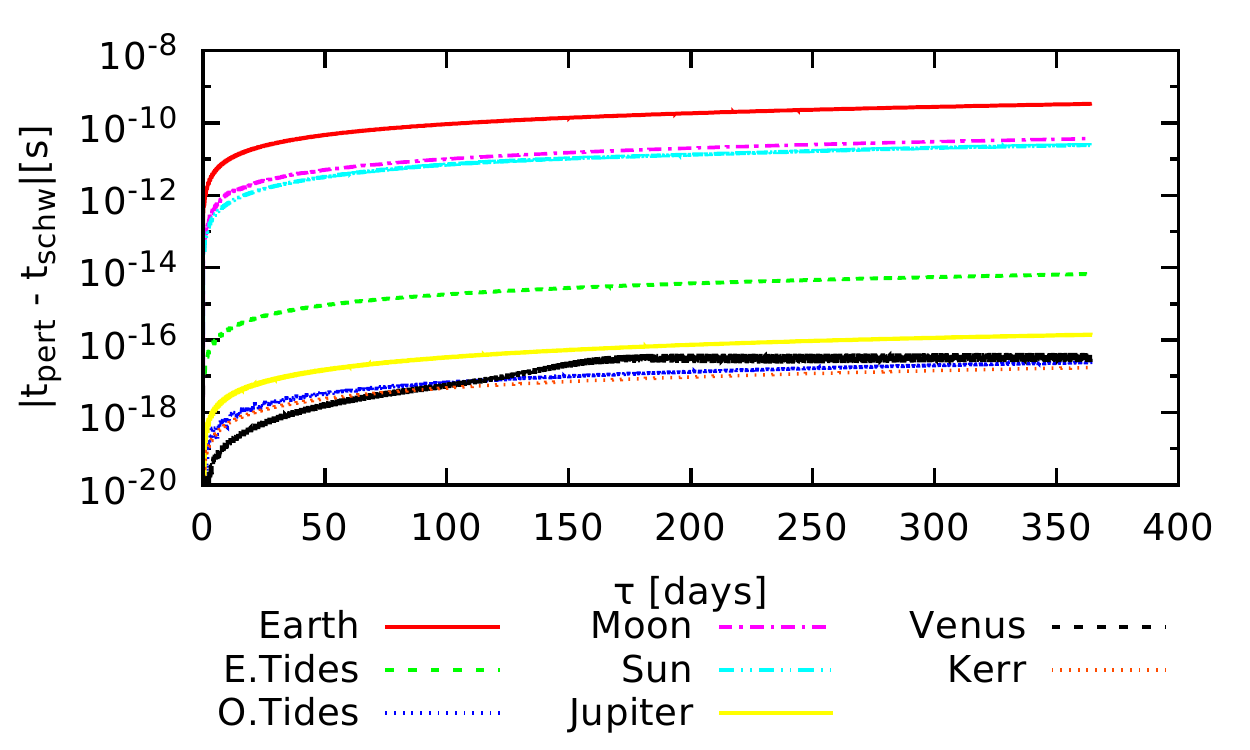}
\caption{The differences in the satellite's position $\Delta L$ (left) and Schwarzschild time $\Delta t$ (right) due to each gravitational perturbation. The time on the abscissa counts days from 1 January 2012 at 7:00 a.m.. The initial values of parameters are: $t_a = 7$ h, $\omega=0^{\circ}$, $\Omega = 0^{\circ}$, $a=29602$ km, $\ecc = 0.007$, $\incl=56^{\circ}$.}
\label{fig:everything_times}
\end{figure}

Variations of the orbital parameters due to gravitational perturbations are very small, therefore, instead of plotting the perturbed orbits, we show the differences between satellite positions on perturbed and unperturbed orbit: \reff{fig:everything_times} left shows differences between positions\footnote{We do not measure the difference in position in length along the orbit, but by 3D distance between both positions: $\Delta L= \vert \vec{r}_{\mathrm perturbed} - \vec{r}_{\mathrm Schwarzschild}\vert$.} and \reff{fig:everything_times} right shows differences in Schwarzschild times, both at the same proper time.
The largest differences in positions $\Delta L$ and times $\Delta t$ are $\sim 10^4$ km and $\sim 0.1 - 1$ ns in one year, and come from the Earth multipoles, while the smallest are $\sim 20$ cm and $\sim 10^{-17}$ s in one year, and come from the Kerr perturbation.

From the two figures it is evident, which perturbations need to be included to reach a desired accuracy of a positioning system, e.g. for a system with accuracy better than $1\ \mathrm{m}$ in one year, all perturbations down to Kerr should be included. For satellites at $r\approx 30.000$ km, the effect of Kerr perturbation is $0.5$ mm per day, therefore it would be in principle sufficient to perform all calculations in double precision. However, because in GNSS we are dealing with quasi-circular orbits, some precision loss can occur due to cancellation effects. Therefore, we performed the calculations in 128-bit floating point numbers, to ensure that Kerr effect is not lost in numerical noise.

\section{Relativistic positioning system}
\label{sec:RPS}
To model the relativistic positioning system in gravitationally perturbed space-time and test its accuracy, we use perturbed satellite orbits from Section \ref{sec:evolution_of_orbits} and simulate a constellation of four satellites orbiting the Earth. The initial orbital parameters of the satellites are: for all satellites $\Omega_i=0\degr$, $a_i= 30000\ \mathrm{km}$, $\ecc = 0.007$, $t_i = 0$ ($i=1,\ldots 4$), for the first two satellites the inclination is $\incl_1 =\incl_2 = 45\degr$ and for the last two it is $\incl_3=\incl_4=135\degr$. The arguments of the apoapsis are $\omega_1 = 270$, $\omega_2 = 315$, $\omega_3 = 275$, $\omega_4 = 320$. The user's coordinates $r_{\rm o}=6371\ \mathrm{km}$, $\theta_{\rm o}=43.97\degr$, $\phi_{\rm o}=14.5\degr$ remain constant during the calculations of its position.\footnote{Note that these are Schwarzschild coordinates and not geographical coordinates on Earth.}

We assume that in real applications of the positioning either the orbital parameters or the satellites' positions would be transmitted to the user as a part of the signal. To account for this in our simulations, we calculated the satellite orbits before starting the positioning. Additionally, we assumed that the position of the user is completely unknown, i.e., we did not start from the last known position.

The satellites' trajectories are parametrized by their proper time $\tau$ and are obtained by numerical integration of \refe{eq:geodesic}. At every time-step of the simulation, each satellite emits a signal and a user on Earth receives signals from all four satellites. The event $\pp_{\rm o} = (t_{\rm o},x_{\rm o},y_{\rm o},z_{\rm o})$ marks the user's Schwarzschild coordinates at the moment of reception of the signals from four satellites. Each satellite emitted a signal at event $\pp_i = (t_i,x_i,y_i,z_i)$, corresponding to $\tau_i$ ($i=1,\ldots,4$). Taking into account that the events $\pp_{\rm o}$ and $\pp_i$ are connected with a light-like geodesic, we calculate $\tau_i$ at the emission point $\pp_i$ using the equation
\beq
  t_{\rm o} - t_i(\tau_i) = T_{\rm f}(\vec{R}_i(\tau_i), \vec{R}_{\rm o})\ ,
\label{eq:positioning}
\eq
where $\vec{R}_i=(x_i,y_i,z_i)$ and $\vec{R}_{\rm o} = (x_{\rm o},y_{\rm o},z_{\rm o})$ are the spatial vectors of the satellites and the user, respectively. The function $T_{\rm f}$ calculates the time-of-flight of photons between $\vec{R}_{\rm o}$ and $\vec{R}_i$ using elliptic integrals and Jacobi elliptic functions as shown in \cite{2005PhRvD..72j4024C} and \cite{vCadevz2010}. The equation \refe{eq:positioning} is actually a system of four equations for four unknown $\tau_i$. Once the values of $\tau_i$ are determined, they define the user's emission coordinates at $\pp_{\rm o} = (\tau_1,\tau_2,\tau_3,\tau_4)$.

To calculate user's position and time in the more customary Schwarzschild coordinates, we use the Schwarzschild coordinates of all four satellites at emission events $\pp_i$, and obtain the first approximation of the user's Schwarzschild coordinates by taking the geometrical approach presented in \cite{vCadevz2010}, which is based on finding the intersection of four spheres with radii $t_{\rm o} - t_i$, centered at $\vec{R}_i=(x_i,y_i,z_i)$. To polish the result, we again solve \refe{eq:positioning}, however, this time it is treated as a system of four equations for four unknown user coordinates, i.e., solving it, gives $(t_{\rm o},x_{\rm o},y_{\rm o},z_{\rm o})$.

The accuracy of this model was tested by comparing the true user coordinates $t_{\rm o}$ and $\vec{R}_{\rm o}$ to the coordinates $t_{\rm o}^{\rm e}$ and $\vec{R}_{\rm o}^{\rm e}$ as determined from emission coordinates. The relative errors, defined as
\beq
\epsilon_t  = \frac{ t_{\rm o} - t_{\rm o}^e}{t_{\rm o}} \hspace{0.7cm}
\epsilon_x = \frac{x_{\rm o} - x_{\rm o}^{\rm e}}{x_{\rm o}} \hspace{0.7cm}
\epsilon_y = \frac{y_{\rm o} - y_{\rm o}^{\rm e}}{y_{\rm o}} \hspace{0.7cm}
\epsilon_z = \frac{z_{\rm o} - z_{\rm o}^{\rm e}}{z_{\rm o}}
\eq
are of the order $10^{-32}-10^{-30}$ for coordinate $t$, and $10^{-28}-10^{-26}$ for $x$, $y$, and $z$. On a laptop\footnote{With the configuration: Intel(R) Core(TM) i7-3610QM CPU @ 2.30GHz, 8GB RAM} the user's position was determined in $0.04\ \mathrm{s}$. By taking into account the last known position of the user, the calculation could be even quicker. These results show that special computing facilites are not required for implementing positioning algorithms within an RPS, i.e. ordinary GNSS receivers are completely adequate for this task.  Because we assumed that the model of the Earth space-time is exact and did not include non-gravitational perturbations, these relative errors reflect the numerical accuracy of the algorithms -- in a real system, the errors would be much higher, depending on the quality of the space-time model.

\section{Conclusions}
\label{sec:conclusions}
We modelled a Relativistic Positioning System by using emission coordinates in perturbed Schwarzschild space-time. 

The gravitational perturbations (Earth multipoles up to $6^{\mathrm{th}}$, the Moon, the Sun, Jupiter, Venus, solid and ocean tides, Kerr effect) were added to the background of the Schwarzschild metric in the weak-field limit with the linear perturbation theory. The solutions (beyond the dominant monopole) were obtained using the RWZ formalism, i.e. the perturbations were expanded in terms of tensor spherical harmonics or normal modes. We used the $c\to\infty$ limit to find connection between our solutions and Newtonian multipole coefficients. The frame-dragging effect of the Earth was taken into account by the first order term in the expansion of the Kerr metric for $a\ll 1$. The result was a perturbed Schwarzschild metric, with perturbations being fully determined by multipole momenta $\overline{M}_{nm}^\oplus$, $\overline{M}_{nm}^\ominus$ and Kerr parameter $a$.

The perturbed metric was used in geodesic equation to obtain satellites' orbits. We investigated the influence of gravitational perturbations on the orbital parameter evolution and on the satellite's position and time. As expected, the biggest effects arise from Earth quadrupole and the Moon, e.g. the differences in satellites' positions are $\Delta L\approx 10^4\ \mathrm{km}$ and times $\Delta t \approx 0.1 - 1\ \mathrm{ns}$ in one year, while the smallest arise from the Kerr effect, e.g. $\Delta L\approx 20\ \mathrm{cm}$ and $\Delta t \approx 10^{-17}$ s in one year. The results for secular evolution of orbital parameters due to Earth quadrupole agree with Newtonian analytical approximations. Note, that we assumed that the model of the metric around Earth is exact, therefore, the reported accuracies reflect only the accuracy of the numerical algorithms.

The perturbed orbits were used to model an RPS and test its accuracy. We find that a user, which receives proper times of four satellites (i.e. the emission coordinates), can determine its position in such RPS with a relative accuracy of the order of $10^{-32}-10^{-30}$ for coordinate $t$, and $10^{-28}-10^{-26}$ for coordinates $x$, $y$, and $z$. On a laptop, it takes $\sim 0.04\ \mathrm{s}$ to determine the user's position with this accuracy, assuming that the user's last position is completely unknown.

Our work shows that general relativity introduces no technical limitations regarding accuracy and speed of calculations if used in positioning systems, while it brings some advantages, e.g. no relativistic corrections are necessary, as relativity is already included in the definition of the positioning system, and clock synchronization is not required -- in fact, it should be omitted, because the system is based on proper times/emission coordinates. Furthermore, because an RPS uses the emission coordinates, which are not tied to the Earth, a GNSS implemented as an RPS would be much more accurate and stable in the long term, provided that the satellite dynamics is known with sufficient accuracy. Consequently, RPS satellites could be used as a stable measure of the proper time and thus serve as clocks with a long term stability.

\section*{Acknowledgments}
We acknowledge the financial support from European Space Agency PECS project {\em Relativistic Global Navigation System}. We thank Pac{\^o}me Delva and Sante Carloni for fruitful discussions.
\appendix
\section{Regge-Wheeler-Zerilli formalism}
\label{app:rwz}
In the RWZ formalism, the metric perturbation $h_{\mu\nu}$ is expanded into a series of independent tensor harmonics, a tensor analog to spherical harmonic functions, labeled by indices $n$ (degree) and $m$ (order). The tensor harmonics contributions with similar properties, i.e. same parity and indices $(n,m)$, are joined together to form independent metric functions, called in \cite{Vishveshwara1970} the normal modes. The full set of these functions represents a complete functional basis for decomposition of metric perturbations, and as such, it is appropriate for solving the linearized Einstein equation.

The general expansion of the metric perturbation $h_{\mu\nu}$ can be written as
\beq
  h_{\mu\nu} = 
  \sum_{n=2}^\infty \sum_{m=-n}^n
  (h_{\mu\nu}^{nm})^{(\rm o)} + (h_{\mu\nu}^{nm})^{(\rm e)} \ ,
  \label{eq:h_serA}
\eeq
where the expansion terms $(h_{\mu\nu}^{nm})^{(\rm o)}$ and $(h_{\mu\nu}^{nm})^{(\rm e)}$ are the odd-parity and the even-parity metric functions (or modes), respectively.  The parity inversion operator $\hat P:\vec{r} \mapsto -\vec{r}$, written in spherical coordinates as $(r,\theta,\varphi)\mapsto (r,\pi-\theta,\varphi+\pi)$, applied to the metric functions yields:
\beq  
  \hat P (h_{\mu\nu}^{nm})^{(\rm e)}
  = (-1)^n  (h_{\mu\nu}^{nm})^{(\rm e)} \>,\qquad
  \hat P (h_{\mu\nu}^{nm})^{(\rm o)}
  = (-1)^{n+1}  (h_{\mu\nu}^{nm})^{(\rm o)}\>.
\eeq
We find it most convenient to work in the gauge from \cite{Regge1957}, where a coordinate transformation ${x'}^{\nu} = x^{\nu} + \xi^\nu$ is proposed, which conserves the background metric and transforms the metric perturbation according to:
\beq
  h_{\mu\nu}' = h_{\mu\nu} - \xi_{\mu;\nu} -\xi_{\nu;\mu} \>,
  \label{eq:trans_metric_pert}
\eeq
in such a way that the resulting metric functions are reduced in complexity. The transformation \cite{Regge1957} also preserves the degree, the order, and the parity, if defined by the metric perturbation. 
In this gauge the even parity metric functions are:
\beq
  (h_{\mu\nu}^{nm})^{(\rm e)} = 
  \left [
  \begin{array}{cc|cc}
  H_0 \chi  & H_1               & 0      & 0                   \\
  \star     & H_2 \chi^{-1}    & 0      & 0                   \\ \hline
   0         & 0                & r^2 K  & 0                   \\
   0         & 0                & 0      & r^2 K \sin^2\theta
  \end{array}
  \right ] Y_n^m \ ,
  \label{eq:even_metric_elemA}
\eeq
and for odd parity, the metric functions are
\beq
  (h_{\mu\nu}^{nm})^{(\rm o)} = 
  \left [
  \begin{array}{cc|cc}
  0    &  0  & -h_0\csc\theta\partial_\varphi  & h_0\sin\theta\partial_\theta  \\
  0    &  0  & -h_1\csc\theta\partial_\varphi  & h_1\sin\theta\partial_\theta  \\ \hline
  \star&\star  & 0  & 0   \\
  \star&\star& 0  & 0
  \end{array}\right] 
  Y_n^m\ ,
  \label{eq:odd_metric_elemA}
\eeq
where $\star$ indicates symmetric part of the tensor, $\chi = 1 - r_{\rm s}/r$, $r_{\rm s}=2G M/c^2$ is the Schwarzschild radius,  and $Y_n^m$ are  spherical harmonics \cite{Abramowitz1964}. Expressions $h_i$, $H_i$, and $K$ depend on Schwarzschild coordinates $(t,r)$ and indices $(n,m)$, which are omitted for clarity. It is shown in \cite{Regge1957} and \cite{Zerilli1970} that in vacuum $H_0 = H_2$, therefore, both functions are marked with $H$.

The linearized Einstein equations for perturbations in the Schwarzschild background preserve pairs of indices $(n,m)$ as well as the parity, and are homogeneous in the case of vacuum. Inserting $h_{\mu\nu}$ from (\ref{eq:h_serA}) into (\ref{equationA}), leads to a set of homogeneous field equations for functions describing each normal mode independently, and are given in \cite[Eqs. C6--C7]{Zerilli1970}. In the following, we calculate their solutions.


From various possible choices of coordinate transformations, we found the Regge-Wheeler gauge the most convenient for the following reasons: (i) the gauge is completely fixed, (ii) the angular and radial dependence are decoupled in the resulting field equations, and (iii) the solutions have a Newtonian limit, which is important when comparing metric tensor elements with their weak-field limits.
%
%
%
\subsection{Time-independent metric perturbations}\label{sec:RWZ_static}
We first consider a stationary space-time case: Schwarzschild background with time-independent perturbations (e.g., a single, non-rotating, slightly non-spherical object) and find solutions of the differential equations from the previous section. 
We treat even and odd parity modes separately.
%
%
\subsubsection{Even parity contributions}
In the case of time-independent perturbations of even parity $H_1=0$ \cite{Regge1957}. It follows that the even metric mode (\ref{eq:even_metric_elemA}) is diagonal:
\beq
 (h_{\mu\nu}^{nm})^{(\rm e)} = 
 {\rm diag} (H\chi, H\chi^{-1}, r^2 K, r^2 K \sin^2\theta) Y_n^m \ .
  \label{eq:even_metric_static}
\eeq
Inserting it in (\ref{equationA}) gives differential equations for functions $H$ and $K$. With substitution
\beq
S(x)= x(x-1)H(x)
\label{SH} 
\eeq
these two functions are determined up to a constant prefactor by \cite{Zerilli1970}:
\begin{align}
  x (1 - x) S''  + (2 x - 1) S'  + w S  &= 0\ , \label{eq:even_M1} \\
  x(x-1)(H-K)' + H & = 0\ ,  \label{eq:even_K1} \\
  2(x-1) H' - (2x-1) K'  - w (H-K) & = 0 \label{eq:even_K2}\ ,
\end{align}
where we use rescaled radius $x = r/r_{\rm s}$, constant $w=(n - 1) (n + 2)$, and derivative ${()}' = \dd/\dd x$. Because we are interested in solutions outside the Schwarzschild radius (i.e., $x>1$), we can in (\ref{eq:even_M1}) use a substitution $x=1/u$ and rewrite it on the domain $u \in [0,1]$:
\beq
  (u-1) u^2 \ddot S + u (3u - 4) \dot S + w S = 0
  \label{eq:even_M_inv}
\eeq
with $\dot{()} = \dd/\dd u$. Because $u=0$ is a regular singular point, we can solve this equation with the Frobenius method \cite{Arfken1985} around $u=0$ and obtain the solution as the superposition of two independent terms:
\beq
  S(u) = A_{nm} u^{n-1} P_n^{(0)}(u) + B_{nm}  u^{-n -2}R_n^{(0)}(u) \ ,
  \label{eq:M_sol}
\eeq
where $A_{nm} $ and $B_{nm} $ are integration constants. The functions $P_n^{(0)}$ and $R_n^{(0)}$ are expressed by Gaussian hypergeometric functions ${}_2F_1$ \cite{Abramowitz1964}:
\begin{align}
P_n^{(0)}(u) &= {}_2F_1 (-1 + n, 1 + n; 2(n+1); u)\\
R_n^{(0)}(u) &= {}_2F_1(-2 -n, -n; -2 n; u).
\end{align}
The first few terms in the Taylor series of $P_n^{(0)}$ and $R_n^{(0)}$ around $u=0$ are
\beq
  \begin{split}
  &P_n^{(0)}(u) = 1 + \frac{1}{2} (n-1) u + 
  \frac{(n+2) n(n-1)}{4 (2n+3)} u^2 \\
  &+ \frac{(n+3)(n^2-1)n}{24(2n+3)} u^3 +
  O(u^4)\>, 
  \end{split}
\eeq
and
\beq
  \begin{split}
  &R_n^{(0)}(u) = 1 - \frac{2+n}{2} u+
  \frac{(n^2-1) (n+2)}{4 (2n-1)}  u^2 \\
  &-\frac{(n^2-4) n (n+1)}{24 (2n-1)} u^3
  +O(u^4)\>.
  \end{split}
\eeq
Using the relation (\ref{SH}) and the equation (\ref{eq:M_sol}), we can write the solution for $H$ as
\beq
  H(r) = 
    A_{nm} \frac{P_n^{(0)}\left(\frac{r_{\rm s}}{r}\right)}{r^n (r-r_{\rm s})} + 
    B_{nm} \frac{r^{n+1} R_n^{(0)}\left(\frac{r_{\rm s}}{r}\right)}{r-r_{\rm s}}  \>.
  \label{eq:H_sol}
\eeq

By combining equations (\ref{eq:even_K1}) and (\ref{eq:even_K2}) we can express function $K$ with $H$:
\beq
  K = H  + \frac{H'}{w} + \frac{(2x-1) H} {w x (x-1)}\>.
  \label{eq:K_formula}       
\eeq
and note that $K$ is fully determined by $H$. Inserting in this expression the solution (\ref{eq:H_sol}) for  $H$, we obtain the solution for $K$:
\beq
  K(r) = A_{nm} r^{-n-1} P_n^{(1)}\left(\frac{r_{\rm s}}{r}\right) + 
         B_{nm} r^n R_n^{(1)}\left(\frac{r_{\rm s}}{r}\right) \ ,
  \label{eq:K_sol}
\eeq
where the functions $P_n^{(1)}$ and $R_n^{(1)}$ are connected to $P_n^{(0)}$ and $R_n^{(0)}$. 
The first few terms of their Taylor expansion around $u=0$ are
\beq
\begin{split}
&P_n^{(1)}(u) = 1
- \frac{(n - 2) (n + 1)}{2 (n-1)}u 
+ \frac{(n-3) n (n+1)}{4 (2 n-1)}u^2 \\
&-\frac{(n-4) (n - 2) n (n+1)}{24 (2 n-1)} u^3
+ O(u^4)\>,
\end{split}
\eeq
and
\beq
\begin{split}
&R_n^{(1)}(u) = 1
+ \frac{n (n+3)}{2 (n + 2)} u 
+ \frac{n (n+1) (n+4)}{4 (2n + 3)} u^2 \\
&+ \frac{n (n +1) (n+3) (n+5)}{24 (2n + 3)} u^3
+ O(u^4)\>.
\end{split}
\eeq

To obtain the complete form of solutions for $H$ and $K$, we need to determine the integration constants $A_{nm}$ and $B_{nm}$ in (\ref{eq:H_sol}) and (\ref{eq:K_sol}).
Let us compare equation (\ref{eq:H_sol}) with its Newtonian counterpart, i.e., the gravitational potential $\Phi$ of a non-rotating object expanded into a series of multipole contributions \cite{Arfken1985}:

\beq
\begin{split}
  \Phi =  \frac{G M}{r} +  \sum_{nm} (M_{nm}^\oplus r^{-n-1} + M_{nm}^\ominus r^n) Y_n^m\ ,
\end{split}
   \label{eq:stat_pot_pert}
\eeq
where $M_{nm}^\oplus$ and $M_{nm}^\ominus$ are time-independent Newtonian spherical multipole momenta and notation $\sum_{nm}\equiv\sum_{n=2}^\infty\sum_{m=-n}^n$ is used.
We choose the sign of (\ref{eq:stat_pot_pert})  so that the force is ${\bf F} = \nabla \Phi$. 
The first term in the sum describes the gravitational potential of the perturbing sources positioned within the radius $r$, while the second term corresponds to those outside $r$. Comparing (\ref{eq:H_sol}) with (\ref{eq:stat_pot_pert}), we notice the same behaviour for $r\gg r_{\rm s}$ (i.e., the superposition of $r^{-n-1}$ and $r^{n}$ functional dependence) in the perturbative part of (\ref{eq:stat_pot_pert}) and it is evident that the coefficients $A_{nm}$ and $B_{nm}$ are related to the multipole momenta. 
The relation between both is found from the weak field approximation
\beq
  \frac{c^2}{2} (1 + g_{00}) \sim \Phi \ .
  \label{eq:lim_g00}
\eeq
By inserting
\beq
  g_{00} = g_{00}^{(0)}  + \sum_{nm}{(h_{00}^{nm})}^{(\rm e)} = \chi \left(-1 + \sum_{nm} H_{nm}\right)
\eeq
into the above relation together with the Newtonian potential (\ref{eq:stat_pot_pert}), we find that in the weak field limit $A_{nm}$ and $B_{nm}$ are asymptotically related to Newtonian spherical multipole momenta $M_{nm}^\oplus$ and $M_{nm}^\ominus$ as
\beq
  A_{nm} \sim \frac{2}{c^2} M_{nm}^\oplus
  \quad {\rm and}\quad
  B_{nm} \sim \frac{2}{c^2} M_{nm}^\ominus \ .
  \label{eq:ab}
\eeq
Note that for finite $c$, $M_{nm}^\oplus$ and $M_{nm}^\ominus$ only approximate $A_{nm}$ and $B_{nm}$, as it is clear from post-Newtonian theory \cite{Kaplan2009}.\footnote{For axial-symmetric case it was shown in \cite{Quevedo1990, Backdahl2005} that the leading orders in the expansion of relativistic (Blanchet-Damour-Thorn) \cite{Damour1991} multipoles  w.r.t. speed of light are identical to Newtonian multipoles.}

%
%
%
\subsubsection{Odd parity contributions}\label{sec:odd_timeindep}
In case of time independent perturbations, the odd metric functions ${(h_{\mu\nu}^{nm})}^{(\rm o)}$ in (\ref{eq:odd_metric_elemA}) have $h_1=0$ \cite{Regge1957}  and can be written with a single function $h_0$ as
\beq
\begin{split}
 (h_{\mu\nu}^{nm})^{(\rm o)} = 
 -h_0 \csc\theta\, {Y_n^m}_{,\varphi}
 (\delta_{0,\mu}\delta_{2,\nu} + \delta_{2,\mu}\delta_{0,\nu}) 
 + h_0 \sin\theta\, {Y_n^m}_{,\theta}
 (\delta_{0,\mu}\delta_{3,\nu} + \delta_{3,\mu}\delta_{0,\nu}) \>.
\end{split}
 \label{eq:odd_metric_static}
\eeq
The function $h_0$ is determined up to a pre-factor by equation 
\beq
  h_0'' + 
  \frac{1}{\chi} 
  \left[ \frac{2}{x^3} - \frac{n(n+1)}{x^2}\right] h_0 = 0 \>,
  \label{eq:odd_h0}
\eeq
We are interested in $h_0$ only at $x>1$ and rewrite this equation using the variable $u=1/x$ on the domain of interest $u\in [0,1]$:
\beq
  (1-u)u^2 \ddot h_0 + 2u(1-u) \dot h_0 + [2 u - n(n+1)] h_0 = 0 \>,
\eeq
The point $u=0$ is a regular singular point, so it can be solved with Frobenius method in a similar way as equation (\ref{eq:even_M_inv}). The solution for $h_0$ is
\beq
  h_0(r) =  
  \alpha_{nm} r^{-n} P_n^{(2)}\left(\frac{r_{\rm s}}{r}\right) + 
  \beta_{nm} r^{n+1} R_n^{(2)}\left(\frac{r_{\rm s}}{r}\right) \>,
  \label{eq:odd_stat_sol}
\eeq
where functions $P_n^{(2)}$ and $R_n^{(2)}$ are:
\begin{align}
P_n^{(2)}(u) &= {}_2F_1(-1+n,2+n;2(n+1);u)\\
R_n^{(2)}(u) &= {}_2F_1(-2-n,1-n;-2n;u)\ .
\end{align}
Their Taylor series around $u= 0$ are
\beq
  \begin{split}
  &P_n^{(2)}(u) =
  1 + \frac{(n - 1) (n + 2)}{2(n+1)} u 
  + \frac{(n - 1) n (n + 2) (n + 3)}{4 (n +1 ) (2n + 3)} u^2\\
  & + \frac{(n - 1) n (n + 3) (n + 4)}{24 (2n + 3)} u^3 +O(u^4)\>,
  \end{split}
\eeq
and
\beq
  \begin{split}
  &R_n^{(2)}(u) =
  1 - \frac{(n-1)(n+2)}{2n} u 
  + \frac{(n^2 -1) (n^2 - 4)}{4 n (2n -1 )} u^2\\
  & -\frac{(n-3) (n^2-4) (n+1)}{24 (2 n-1)}u^3+ O(u^4)\>.
  \end{split}
\eeq

To determine the constants $\alpha_{nm}$ and $\beta_{nm}$ in (\ref{eq:odd_stat_sol}), 
we note that off-diagonal terms in the metric tensor are associated with frame-dragging effects. 
Since we are working in a weak field limit, we consider only the frame-dragging effect of the Earth, and neglect frame-dragging effects arising from other objects. Consequently, we set $\beta_{nm} = 0$, because the second term in (\ref{eq:odd_stat_sol}) is due to objects outside $r$.

To determine $\alpha_{nm}$, we notice that for $n=1$ and $m=0$ the corresponding $h_0$ matches the weak field and slow rotation approximation of the Kerr metric: if $r \gg r_{\rm s}$ and angular parameter of the central object is $a \ll 1$, then for $\alpha_{10} =  a r_{\rm s} \sqrt{4\pi/3}$ it follows
\beq
  h_0(r) = a \frac{r_{\rm s}}{r} \sqrt{\frac{4\pi}{3}}\ ,
  \label{eq:h0_stat_sol_kerr}
\eeq
where we keep only the terms linear in $a$.

For higher multipoles ($n>1$), it turns out that their dependence on $a$ is not linear \cite{Hartle1967a}. Therefore, the only multipole we include in the odd-parity metric function is the monopole, i.e., the one belonging to the linear (in $a$) part of the Kerr effect.
%
%
%
\subsection{Time-dependent metric perturbations}\label{sec:timedep}
In this section we consider a slowly rotating Earth, which is also under a weak gravitational influence of other nearby moving objects. 
Therefore, we study time dependent perturbations of the Schwarzschild metric around the Earth, which is slowly rotating around $z$ axis with angular velocity $\Omega$. Due to the Earth's rotation its multipoles vary periodically. Earth's tides introduce additional time dependency in its multipoles (additional variability with different frequency, phase and varying amplitude, depending on the position of the Moon and the Sun). In addition, the gravitational influence of other objects introduces time dependent perturbations to the space-time around the Earth, because their relative positions change with time. These perturbations can be expanded in a series of multipoles and treated with the same procedure as the Earth's multipoles.
We consider time dependent metric perturbations for the case of perturbations oscillating slowly with angular velocities $\omega$, which are smaller or of the same order of magnitude as $\Omega$.
All angular velocities are defined with respect to the Schwarzschild time $t$. 
\subsubsection{Even-parity contributions}\label{sec:even_timedep}
Even-parity modes $(h_{\mu\nu}^{nm})^{(\rm e)}$ in (\ref{eq:even_metric_elemA}) are connected to the Newtonian gravitational potential $\Phi$, which in the case of  time dependent multipoles can be written as:
\beq
  \Phi
   = \frac{G M}{r}
       + 
  \sum_{nm} (M_{nm}^\oplus (T)r^{-n-1} + M_{nm}^\ominus (T) r^n) Y_n^m  \>,
  \label{eq:pot_time_dep}
\eeq
where $T=ct$. 
Alternatively, it can be written in frequency domain as:
\beq
  \Phi  =
  \frac{G M}{r} + 
             \sum_{nm} \int_{-\infty}^\infty  {\rm d} k \, e^{\ii kT} \times
             \left[\widetilde M_{nm}^\oplus (k) r^{-n-1} + \widetilde M_{nm}^\ominus (k) r^n\right] Y_n^m  \>,
\eeq
where $k$ is the wavenumber and $\widetilde M_{nm}^\oplus$, $\widetilde M_{nm}^\ominus$ are the Fourier transforms of time dependent multipoles: 
\beq
  \widetilde M_{nm}^v (k)=  \frac{1}{2\pi}\int_{-\infty}^\infty  {\rm d} T \, e^{-\ii k T}  {M}_{nm}^v (T) \,\>,
\eeq
where $v=\oplus,\ominus$.

Each time dependent multipole generates a time-dependent even metric perturbation ${(h_{\mu\nu}^{nm})}^{(\rm e)}$. 
Functions $H$, $H_1$, and $K$ determining the modes can be expressed with their Fourier transforms:
\beq
 \left(H(T, r), H_1(T, r), K(T, r)\right)= 
   \int_{-\infty}^\infty  {\rm d} k \, e^{\ii k T} (\widetilde H(k, r),\widetilde H_1(k, r),\widetilde K(k, r)) \>.
  \label{eq:even_timedep}
\eeq

Using ansatz (\ref{eq:even_timedep}) in field equations (C7) from \cite{Zerilli1970} 
yields only three independent differential equations (with $()' = \dd /\dd r$) \cite{Regge1957}:
\begin{align}
  \ii k \left( \widetilde K' + \frac{\widetilde K-\widetilde H}{r}  - \frac{r_{\rm s}}{2r^2 \chi}\widetilde K\right) -
  \frac{q} {r^2} \widetilde H_1  &= 0 \label{eq:time_even1}\\
  (\chi \widetilde H_1)' - \ii k(\widetilde K+\widetilde H) &= 0 \label{eq:time_even2}\\
  \ii k \widetilde H_1 + \chi (\widetilde K-\widetilde H)' - \frac{r_{\rm s}}{r} \widetilde H &= 0\ ,
  \label{eq:time_even3}
\end{align}
and an algebraic relation \cite{Zerilli1970}:
\beq
   \left[\frac{3 r_{\rm s}}{r} + w\right] \widetilde H
   +\ii \left[2kr - q \frac{r_{\rm s}}{2 k r^2} \right] \widetilde H_1
   -\left[w + \frac{r_{\rm s}}{r} - \frac{2}{\chi} 
           \left(\frac{r_{\rm s}^2}{(2 r)^2}+ (kr)^2 \right) \right] \widetilde K = 0 \ ,
 \label{eq:time_alg}
\eeq
where $q=n(n+1)$. With variables $x=r/r_{\rm s}$ and $\kappa = k r_{\rm s}$, we can write this algebraic relation in a dimensionless form
\beq
  \left[\frac{3}{x} + w \right] \widetilde H + 
  \ii\left[2 \kappa x - \frac{q}{2 \kappa x^2} \right] \widetilde H_1
  -\left[w + \frac{1}{x} - \frac{2x}{x - 1} \left(\frac{1}{(2x)^2} + (\kappa x)^2\right)\right] \widetilde K = 0 \>. 
\label{eq:time_alg_dimless}
\eeq

Because in our studies $\kappa \ll 1$, we solve equations (\ref{eq:time_even1}) 
- (\ref{eq:time_alg}) perturbatively in $\kappa$. We assume that $\widetilde H$, $\widetilde H_1$, and $\widetilde K$ are smooth functions of $\kappa$, and write them as a power series of $\kappa$. 
We find that an appropriate expansion of these functions for  $\kappa \rightarrow 0$ has the form
\beq
  (\widetilde H,\widetilde H_1,\widetilde K) 
  \sim \widetilde N(\kappa) \sum_{i=0}^\infty \kappa^{2i} 
  (\widetilde a_i( r), \ii \kappa \widetilde b_i( r), \widetilde c_i( r))\>.
  \label{eq:k_exp}
\eeq

Inserting this ansatz in the equations (\ref{eq:time_even1})-(\ref{eq:time_even3}) and neglecting all higher than leading terms in the expansion (\ref{eq:k_exp}), gives us two solutions for each function. 
The leading orders of $\widetilde{H}$ and $\widetilde{K}$ are given in (\ref{eq:H_sol}) and (\ref{eq:K_sol}), respectively, where instead of $A_{nm}, B_{nm}$ from (\ref{eq:ab}) we use another set of constants $\widetilde A_{nm}, \widetilde B_{nm}$ to describe a general case:

\beq
  \widetilde H(k, r) \sim 
    \widetilde A_{nm} (k) \frac{P_n^{(0)}\left(\frac{r_{\rm s}}{r}\right)}{r^n (r-r_{\rm s})} + 
    \widetilde B_{nm} (k) \frac{r^{n+1} R_n^{(0)}\left(\frac{r_{\rm s}}{r}\right)}{r-r_{\rm s}}  + O(\kappa^2) \>.
  \label{eq:H_tilde}
\eeq

\beq
  \widetilde K(k, r) \sim \widetilde A_{nm} (k) r^{-n-1} P_n^{(1)}\left(\frac{r_{\rm s}}{r}\right) + 
         \widetilde B_{nm} (k) r^n R_n^{(1)}\left(\frac{r_{\rm s}}{r}\right) + O(\kappa^2) \>.
  \label{eq:K_tilde}
\eeq
From the algebraic relation  (\ref{eq:time_alg}) we get the leading orders of $\widetilde H_1$:
\beq
  \widetilde H_1 (k, r) \sim 
  - \frac{\\i \kappa}{q}\left(\frac{r}{r_{\rm s}}\right)^2 \left [ \frac{6 r_{\rm s}}{r}+ w (\widetilde H+\widetilde K) - \frac{4 \widetilde K r_{\rm s}}{r-r_{\rm s}} \right] + O(\kappa^3) \>.
  \label{eq:H1_per_tilde}
\eeq
and using the above solutions for $\widetilde H$ and $\widetilde K$, (\ref{eq:H_tilde}) - (\ref{eq:K_tilde}), we can rewrite this in a more explicit form:
\beq
  \widetilde H_1(k, r)
   \sim \widetilde A_{nm}(k)  
      \frac{r^{-n+1} P_n^{(3)}\left(\frac{r_{\rm s}}{r}\right)}{r_{\rm s} (r-r_{\rm s})}   + 
      \widetilde B_{nm}(k)
      \frac{r^{n+2} R_n^{(3)}\left(\frac{r_{\rm s}}{r}\right)}{r_{\rm s} (r-r_{\rm s})} \>,
   \label{eq:H10_sol_explicit}
\eeq
where functions $P_n^{(3)}$ and $R_n^{(3)}$ are given as a series in $u=r_{\rm s}/r$ for $u\to 0$:
\beq
\begin{split}
  &P_n^{(3)}(u) =
   \frac{2}{n}+\frac{n^2+3 n+1}{(n+1) (n+2)}u 
  + \frac{n^3+5 n^2+6 n+3}{2 (n+2)(2 n+3)}u^2\\ &+ 
  \frac{n^3 + 6 n^2+8 n+6}{12 (2 n+3)}u^3 + O(u^4)\>,
\end{split}
\eeq
and
\beq
\begin{split}
  &R_n^{(3)}(u) =
   -\frac{2}{n+1}+\frac{n^2-n-1}{(n-1) n}u
  -\frac{n^3-2 n^2-n-1}{2 (n-1) (2 n-1)}u^2\\ &+ \frac{n^3-3 n^2-n-3}{12 (2 n-1)}u^3 + O(u^4)\ .
\end{split}
\eeq
By considering the weak field limit (\ref{eq:lim_g00}) we find that 
\beq
  \widetilde A_{nm} \sim  \frac{2}{c^2} \widetilde M_{nm}^\oplus
  \quad{\rm and}\quad 
  \widetilde B_{nm} \sim  \frac{2}{c^2} \widetilde M_{nm}^\ominus\>.
\eeq

The metric perturbation expressed with these functions is accurate up to the linear order in frequency. Since higher order perturbations naturally give rise to contributions with higher orders of frequency, our approximation of the perturbation is consistently linear, i.e., it is linear in frequency and in the order of perturbation.
%
%
%
\subsubsection{Odd-parity contributions}
For odd-parity contribution to the metric ${(h_{\mu\nu}^{nm})}^{(\rm o)}$ (\ref{eq:even_metric_elemA}) we use the same notation for solutions $h_0, h_1$ as in (\ref{eq:even_timedep}):
\beq
  (h_0(T,r), h_1(T,r))=\int_{-\infty}^\infty  {\rm d} k \, e^{\ii k T} (\tilde h_0(k,r), \tilde h_1(k,r)) \>.
 \label{eq:odd_timedep}
\eeq
Because we are interested only in persistent phenomena, we limit ourselves to real wavenumbers, $k\in\bR$. A more detailed discussion of all possible solutions is given in e.g., \cite{Vishveshwara1970}.
Using ansatz (\ref{eq:odd_timedep}) in field equations (C6) from \cite{Zerilli1970} 
 for odd metric functions, we obtain only two independent equations
\begin{align}
 \ii k \tilde h_0 - \chi (\chi \tilde h_1)' &= 0\>, \label{eq:h_0} \\ 
  k^2 \tilde h_1 + \ii k (\tilde h_0'- 2\frac{\tilde h_0}{r}) - w \chi \frac{\tilde h_1}{r^2} &=0\>,
\end{align}
with $()' = \dd /\dd r$.
It was shown in \cite{Regge1957} that we can use a substitution for $h_1$:
\beq
  Q = \chi \frac{\tilde h_1}{r}
  \label{eq:Q}
\eeq
to eliminate $h_0$ from both equations. Thereby we obtain a wave equation for $Q$ in the form
\beq
  \frac{\dd^2}{\dd r_*^2} Q + k_{\rm eff}^2 Q = 0 \>,
  \label{eq:wave_eq_Q}
\eeq
where $r_*$ is modified radius defined as 
\beq
  \dd r_* = \chi^{-1} \dd r
  \quad{\rm or}\quad 
  r^* = r + r_{\rm s} \log(r-r_{\rm s}) + {\rm const.}
\eeq 
and an effective wave number 
\beq
  k_{\rm eff}^2  = k^2 - \frac{n(n+1)\chi}{r^2}  + 3 \frac{r_{\rm s} \chi}{r^3} \>.
\eeq
By knowing $Q$, we can express $\tilde h_1$
from (\ref{eq:Q}) and write $\tilde h_0$ using (\ref{eq:h_0}) as:
\beq
\tilde h_0 = -\frac{\ii}{k} \chi (r Q)'  \>.
\eeq

From equation (\ref{eq:wave_eq_Q}) we find that in the limit $r\to\infty$ its solution is $Q(r)\asymp \sin (k r + \phi)$. This determines asymptotic behavior of $\tilde  h_0$ and $\tilde h_1$:
\beq
  \tilde h_1(r) \asymp r \sin (k r + \phi)
  \quad\textrm{and}\quad
  \tilde h_0(r) \asymp r \cos(kr + \phi)\>.
  \label{eq:h0h1_timedep}
\eeq
We see that the asymptotic behaviour of solutions $\tilde  h_0$ and $\tilde h_1$ is not flat. Because we neglect frame-dragging effects arising from other objects, as mentioned in \ref{sec:odd_timeindep}, we neglect time dependent odd-parity contributions all together.

\section{Multipole momenta}\label{app:multipoles}
\subsection{Planets, Moon, Sun}
%
%
%
\label{sec:mult_planets}
%
%
We consider the Newtonian gravitational potential of celestial bodies $\Phi^\ominus$ acting on a point of space around the Earth ${\bf r}$ in the Earth-centered inertial (ECI) system, such as International Celestial Reference System (ICRS)/J2000.0 \cite{Ma1997, Souchay2006, Petit2010}, written as
\beq
  \Phi^\ominus(t, {\bf r})  = \sum_i \frac{G M_i}{\|{\bf r}_i-{\bf r}\|}\>, 
  \label{eq:pert_planets} 
\eeq
where $M_i$ are the masses of celestial bodies and ${\bf r}_i$ are vectors pointing to the celestial body. The latter are functions of time and are supplied by JPL's Horizons System \cite{Horizons2013} in ICRS/J2000.0. 


The potential $\Phi^\ominus$ \refe{eq:pert_planets} can be expanded in terms of spherical harmonics
\beq
  \Phi^\ominus(t, {\bf r}) 
    = \sum_{n=0}^\infty \sum_{m=-n}^n 
    r^n M_{nm}^\ominus(t) Y_n^m(\theta,\varphi)\>,
  \label{eq:U_celest}
\eeq
where we identify the interior multipole moments associated to the celestial bodies given by
\beq
  M_{nm}^{\ominus}(t) = 
  \frac{4\pi G}{2n+1}
  \sum_i\frac{M_i}{r_i^{n+1}(t)} {Y_n^m}^*(\theta_i(t),\varphi_i(t)) \>,
\eeq
which are time dependent. Note that $M_{nm}^\ominus$ are complex with the symmetry 
\beq
  {M_{nm}^\ominus}^* = (-1)^m M_{n,-m}^\ominus \>,
\eeq
which makes the potential $\Phi^\ominus$ a real valued function.

\subsection{Earth tides}
%
%
%
\def\Rx{{\bf R}_{\rm x}}
\def\Ry{{\bf R}_{\rm y}}
\def\Rz{{\bf R}_{\rm z}}
\label{sec:mult_tides}
The Earth's Newtonian gravitational potential $\Phi^\oplus$, called the geopotenial, is time varying due to gravitational influence of the Moon, the Sun, and rotation of the deformed Earth. 


We start with the Newtonian gravitational potential $\Phi$ in the Earth centered Earth fixed (ECEF) system, such as International Terrestrial Reference Systems (ITRS) \cite{Petit2010}, and then transform it into Earth-centered inertial (ECI) system, such as International Celestial Reference System (ICRS)/J2000.0 \cite{Ma1997, Souchay2006, Petit2010}, for use in satellite dynamics calculations. 

In the spherical coordinates based on ITRS, the geopotential can be written in terms of the spherical harmonics $Y_n^m$ as
\beq
  \Phi^\oplus(t, {\bf r}) = 
  \sum_{n=0}^\infty \sum_{m=-n}^n \frac{M_{nm}(t)}{r^{n+1}} Y_n^m(\theta,\varphi) \ ,
  \label{eq:newt_pot}
\eeq
where $M_{nm}(t)\in \bC$ are the complex multipole moments depending on time $t$ and possessing the symmetry $M_{n,-m} = (-1)^m M_{n,m}^*$. In the chosen coordinate system there is no dipole contribution: $M_{1m}=0$. 

The geopotential is expressed using the time dependent normalized geopotential coefficients $\overline{C}_{nm}(t)$ and $\overline{S}_{nm}(t)$ and normalized Legendre polynomial $P_{nm}$ as \cite{Torge2001}
\beq
  \begin{split}
  & \Phi^\oplus (t, {\bf r})
  = \frac{G M_\Earth}{r} \bigg\{1 + \sum_{n=2}^\infty  
  \left (\frac{r_\Earth}{r} \right)^n\cdot \\
  &\sum_{m=0}^n 
  [ \overline{C}_{nm}(t) \cos (m \varphi) + \overline{S}_{nm}(t) \sin(m \varphi)]
  \overline{P}_{nm}(\cos \theta)
  \bigg\}\>,
  \end{split}
  \label{eq:newt_pot_geodesy_normalized}
\eeq
where $M_\Earth$ and $r_\Earth$ are the mass and the mean radius of Earth, respectively. The normalized Legendre polynomial are defined as
\beq
  \overline{P}_{nm} = (-1)^m N_{nm} P_n^m \ ,
\eeq
where the normalization factor is
\beq
  N_{nm} = \sqrt{\frac{(2-\delta_{m,0})(2n+1)(n-m)!}{(n+m)!}} \>,
\eeq
and $P_n^m$ are standard Legendre polynomials \cite{Abramowitz1964}.

The sum over $n$ in \refe{eq:newt_pot_geodesy_normalized} runs only from the quadrupole term due to choice of coordinate system, and by definition, $S_{n0}=0$ for all $n$. 

The normalized geopotential coefficients are connected to the complex multipoles in (\ref{eq:newt_pot}) for positive orders $m>0$ via formula
\beq
  M_{nm} = 
  (-1)^m\sqrt{\frac{4\pi}{2-\delta_{m,0}}}  G M_\Earth  r_\Earth^n T_{nm} \>,
  \label{eq:T_to_M}
\eeq
where we introduce a complex normalized geopotential coefficient 
\beq
  T_{nm} = \overline{C}_{nm} - \ii\, \overline{S}_{nm} \>,
  \label{eq:comp_geo_coef}
\eeq
which is frequently used in tide calculations as it allows us to work with both real geopotential coefficients in the same expression. 

The complex normalized geopotential coefficients $T_{nm}$ can be decomposed into a sum of constant (time average) coefficients $T_{nm}^0$ and its perturbation, i.e., time dependent coefficients $T_{nm}^{\rm e}(t)$ and $T_{nm}^{\rm o}(t)$ corresponding to Earth and ocean tides, respectively:
\beq
  T_{nm}(t) = T_{nm}^0 + T_{nm}^{\rm e}(t) + T_{nm}^{\rm o}(t) \>.
\eeq
The constant contributions $T_{nm}^0$ are up to degree 360 available from Earth Gravitational Model 1996 (EGM96) \cite{Lemoine1998}.  In the following, we separately discuss each of the perturbations to $T_{nm}$.
\subsubsection{Solid Earth tides}
Following International Earth Rotation and Reference Systems service (IERS) Conventions \cite[Ch. 6]{Petit2010} and \cite[Sec. 3.7.2]{Montenbruck2005}, we write the complex geopotential coefficients of solid Earth as a sum of contributions due to presence of the Moon ($j=1$) and the Sun ($j=2$):
\beq
  T_{nm}^{\rm e}(t) 
   =\frac{k_{nm}}{2n+1}
  \sum_{j=1}^2 \frac{M_j}{M_\Earth}
  \left(\frac{r_\Earth}{r_j}\right)^{n+1} 
  \overline{P}_{nm} (\sin\theta_j ) e^{-\ii m \varphi_j} \>,
  \label{eq:tides_solid}
\eeq
where $M_j$, $r_j$, $\theta_j$ and $\varphi_j$ are the mass, the distance from the Earth center, the latitude and the east longitude from Greenwich, respectively, of the $j$th body in ITRS. The position $(r_j, \theta_j, \varphi_j )$ as a function of time is provided by Astronomical Almanac \cite{Almanac2012} together with The Explanatory Supplement \cite{Seidelmann1992}. 

The $k_{nm}$ are the nominal Love numbers describing Earth's response to $(n,m)$-multipoles of the external potential and depend on the considered Earth model. Here we assume that the Earth is elastic and use the numbers from \cite[Table 6.3]{Petit2010}. 
%
With such description, we capture only the main body deformations, whereby the motion of the poles is left out of discussion. 
The model describing perturbations of multipole coefficients is given in \cite[Sec. 5.2.8.]{Moyer2005} and concrete data can be found in \cite[Sec 6.4.]{Petit2010}. 
%
%
%
\subsubsection{Ocean tides}
%

The ocean tide can be broken down into its independent constituents representing perpetual dynamics of ocean surface of incommensurable frequency. 
%
%
Few of the most influential tide constituents used in our analysis are (marked with Darwin's symbols): principal lunar semidiurnal ($M_2$), principal solar semidiurnal ($S_2$), larger lunar elliptic semidiurnal ($N_2$), and lunar diurnal ($K_1$ and $O_1$). 
%
%
For a given constituent $f$, labeled by the Doodson number $m_i$, we introduce Doodson coefficients $n_i$,
\beq 
  m_1 = n_1\>,\qquad m_i = n_i + 5\quad{\rm for}\quad i=2,\ldots,6 \>. 
\eeq
%
According to \cite[Sec. 3.7.2]{Montenbruck2005} and IERS Conventions \cite{Petit2010}, the geopotential coefficients due to ocean tides can be represented as a sum over constituents $f$
\beq
   T_{nm}^{\rm o}(t)
   = \sum_f \sum_{s \in \{+,-\}} 
   [{\cal C}_{f,nm}^s -\ii s\, {\cal S}_{f,nm}^s] e^{s\,\ii \theta_f(t)} \ ,
  \label{eq:tides_ocean}
\eeq
where $\theta_f(t)$ is the Doodson argument defined as a linear combination of six Doodson fundamental arguments $\beta_i$:
\beq
  \theta_f(t) = \sum_{i=1}^6 n_i \beta_i(t) \>,
\eeq
and ${\cal C}_{f,nm}^s$ and ${\cal S}_{f,nm}^s$ are geopotential harmonic amplitudes corresponding to constituent $f$ and multipoles indices $(n,m)$. 
The amplitudes  ${\cal C}_{f,nm}^s$ and ${\cal S}_{f,nm}^s$ based on Finite element solutions of global tides for year 2004 (FES2004) \cite{Lyard2006} are provided by R. Biancale \cite{Biancale2012}. 
\section*{References}
\bibliography{biblio}

\providecommand{\newblock}{}
\begin{thebibliography}{10}
\expandafter\ifx\csname url\endcsname\relax
  \def\url#1{{\tt #1}}\fi
\expandafter\ifx\csname urlprefix\endcsname\relax\def\urlprefix{URL }\fi
\providecommand{\eprint}[2][]{\url{#2}}

\bibitem{Ashby2003}
Ashby N 2003 {\em Living Reviews in Relativity\/} {\bf 6} 43

\bibitem{Pascual-Sanchez2007}
Pascual-S\'{a}nchez J~F 2007 {\em Ann. Phys. (Leipzig)\/} {\bf 16} 258--273

\bibitem{Coll1991}
Coll B and Morales J~A 1991 {\em Journal of Mathematical Physics\/} {\bf 32}
  2450 ISSN 00222488

\bibitem{Rovelli2002}
{Rovelli} C 2002 {\em Phys. Rev. D\/} {\bf 65}(4) 044017

\bibitem{Blagojevic2002}
{Blagojevi{\'c}} M, {Garecki} J, {Hehl} F~W and {Obukhov} Y~N 2002 {\em Phys.
  Rev. D\/} {\bf 65} 044018

\bibitem{Coll2003}
Coll B 2003 {A principal positioning system for the Earth} {\em Journ{\'{e}}es
  2002 - syst{\`{e}}mes de r{\'{e}}f{\'{e}}rence spatio-temporels. Astrometry
  from ground and from space, Bucharest, 25 - 28 September 2002, edited by N.
  Capitaine and M. Stavinschi, Bucharest: Astronomical Institute of the
  Romanian Academy, Paris: Observato\/} vol~14 ed Capitaine N and Stavinschi M
  pp 34--38

\bibitem{Tarantola2009}
{Tarantola} A, {Klimes} L, {Pozo} J~M and {Coll} B 2009 {\em ArXiv e-prints\/}
  (\textit{Preprint} \eprint{0905.3798})

\bibitem{vCadevz2010}
{\v C}ade{\v{z}} A, Kosti\'{c} U and Delva P 2010 Mapping the spacetime metric
  with a global navigation satellite system, european space agency, the
  advanced concepts team, ariadna final report (09/1301) Tech. rep. European
  Space Agency

\bibitem{2005PhRvD..72l4016F}
{Fang} H and {Lovelace} G 2005 {\em \prd\/} {\bf 72} 124016

\bibitem{2005PhRvD..71j4003M}
{Martel} K and {Poisson} E 2005 {\em \prd\/} {\bf 71} 104003

\bibitem{Regge1957}
Regge T and Wheeler J~A 1957 {\em Phys. Rev.\/} {\bf 108}(4) 1063--1069

\bibitem{Zerilli1970}
Zerilli F~J 1970 {\em Phys. Rev. D\/} {\bf 2}(10) 2141--2160 see erratum by
  Zerilli in Appendix A-7 of \cite{Gurzadyan2005}

\bibitem{Quevedo1990}
Quevedo H 1990 {\em Fortschritte der Physik/Progress of Physics\/} {\bf 38}
  733–840

\bibitem{Backdahl2005}
Backdahl T and Herberthson M 2005 {\em Classical and Quantum Gravity\/} {\bf
  22} 3585

\bibitem{Hartle1967}
Hartle J~B 1967 {\em Astrophysical Journal\/} {\bf 150} 1005

\bibitem{Hartle1967a}
Hartle J~B and Sharp D~H 1967 {\em Astrophysical Journal\/} {\bf 147} 317

\bibitem{Horizons2013}
{NASA JPL's Solar System Dynamics group} 2013 {HORIZONS system Web-Interface}
  \urlprefix\url{http://ssd.jpl.nasa.gov/horizons.cgi}

\bibitem{Blitzer1970}
Blitzer L 1970 Astronautics 453 -- handbook of orbital perturbations Tech. rep.
  University of Arizona

\bibitem{2005PhRvD..72j4024C}
{{\v C}ade{\v z}} A and {Kosti{\' c}} U 2005 {\em Phys. Rev. D\/} {\bf 72}
  104024

\bibitem{Vishveshwara1970}
Vishveshwara C~V 1970 {\em Phys. Rev. D\/} {\bf 1}(10) 2870--2879

\bibitem{Abramowitz1964}
Abramowitz M and Stegun I~A 1964 {\em Handbook of Mathematical Functions\/} 5th
  ed (New York: Dover)

\bibitem{Arfken1985}
Arfken G 1985 {\em Mathematical Methods for Physicists\/} 3rd ed (Orlando:
  Academic Press)

\bibitem{Kaplan2009}
Kaplan J~D, Nichols D~A and Thorne K~S 2009 {\em Phys. Rev. D\/} {\bf 80}(12)
  124014

\bibitem{Damour1991}
Damour T, Soffel M and Xu C 1991 {\em Phys. Rev. D\/} {\bf 43}(10) 3273--3307

\bibitem{Ma1997}
Ma C, Feissel M and Service I~E~R 1997 {\em Definition and Realization of the
  International Celestial Reference System by {VLBI} Astrometry of
  Extragalactic Objects\/} IERS technical note (Central Bureau of IERS,
  Observatoire de Paris)

\bibitem{Souchay2006}
Souchay J 2006 {\em The International Celestial Reference System and Frame:
  ICRS Center Report for 2001-2004\/} IERS technical note (Verlag des
  Bundesamtes f{\"u}r Kartographie und Geod{\"a}sie) ISBN 9783898888028

\bibitem{Petit2010}
Petit G and Luzum B 2010 {IERS Conventions, IERS Technical Note No. 36} Tech.
  rep. International Earth Rotation and Reference Systems Service

\bibitem{Torge2001}
Torge W 2001 {\em Geodesy\/} (Walter de Gruyter)

\bibitem{Lemoine1998}
Lemoine F and Center G~S~F 1998 {\em Geopotential Model EGM96\/} NASA technical
  paper (National Aeronautics and Space Administration, Goddard Space Flight
  Center)

\bibitem{Montenbruck2005}
Montenbruck O and Gill E 2005 {\em Satellite Orbits: Models, Methods,
  Applications\/} (Springer Verlag)

\bibitem{Almanac2012}
{US Nautical Almanac Office} 2012 {\em Astronomical Almanac for the Year 2013
  and Its Companion, the Astronomical Almanac Online\/} Astronomical Almanac
  For The Year ({U.S. Government Printing Office}) ISBN 9780707741284

\bibitem{Seidelmann1992}
Seidelmann P~K, Britain G and Observatory U~S~N 1992 {\em Explanatory
  supplement to the astronomical almanac\/} [rev. ed.]. ed (University Science
  Books, Mill Valley, Calif. :) ISBN 0935702687

\bibitem{Moyer2005}
Moyer T 2005 {\em Formulation for Observed and Computed Values of Deep Space
  Network Data Types for Navigation\/} JPL Deep-Space Communications and
  Navigation Series (John Wiley \& Sons) ISBN 9780471726173

\bibitem{Lyard2006}
{Lyard} F, {Lefevre} F, {Letellier} T and {Francis} O 2006 {\em Ocean
  Dynamics\/} {\bf 56} 394--415

\bibitem{Biancale2012}
Biancale R 2012 Updates to the iers conventions (2010): Chapter 6 geopotential:
  Harmonic coefficients of the main waves of fes2004 accessed: 10. September
  2013

\bibitem{Gurzadyan2005}
Gurzadyan V, Makino J, Rees M~J, Meylan G, Ruffini R and Wheeler J~A 2005 {\em
  Black Holes, Gravitational Waves and Cosmology\/} Advances in Astronomy and
  Astrophysics Series (Cambridge Scientific Publishers Limited) ISBN
  9781904868255

\end{thebibliography}
%
%
%
\end{document}